\documentclass{aa}

\bibpunct{(}{)}{;}{a}{}{,} 

\usepackage{graphicx,color,epstopdf,hyperref}
\usepackage{soul}
\usepackage[varg]{txfonts}
\newcommand{\mstar}{M$_{\rm *}$}

\newcommand{\msun}{M$_\odot$}

\begin{document}

   \title{Diversity of dwarf galaxy IR-submm emission patterns: CLUES from hydrodynamical simulations}


   \author{Isabel M.E. Santos-Santos
          \inst{\ref{inst1},\ref{inst2}}\thanks{isabelm.santos@uam.es},         
          Rosa Dom\'inguez-Tenreiro\inst{\ref{inst1},\ref{inst2}}, Gian Luigi Granato\inst{3}, Chris B. Brook\inst{4,5}, Aura Obreja\inst{6}\\
          }
\institute{   Depto. de F\'isica Te\'orica, Universidad Aut\'onoma de Madrid, E-28049 Cantoblanco, Madrid, Spain\label{inst1}
 \and Astro-UAM, UAM, Unidad Asociada CSIC\label{inst2}
\and  Osservatorio Astronomico di Trieste, INAF, Via Tiepolo 11, I-34131 Trieste, Italy
\and Instituto de Astrof\'isica de Canarias, C/ V\'ia L\'actea s/n, 38205 La Laguna, Tenerife, Spain
\and Depto. de Astrof\'isica, Universidad de La Laguna, Av. del Astrof\'isico Francisco S\'anchez s/n, 38206 La Laguna, Tenerife, Spain
\and New York University Abu Dhabi, PO Box 129188, Saadiyat Island, Abu Dhabi, United Arab Emirates
}

   \date{Received ; accepted }

 
  \abstract
  {
The spectral energy distributions (SEDs) of low-mass low-metallicity (dwarf) galaxies  are a challenging piece of the puzzle of galaxy formation in the near Universe. These SEDs show some particular features in the submillimeter to far-infrared wavelength range compared to normal larger galaxies that
cannot be explained by the current models.
 } 
  {We aim to explain the particular
emission features of low-mass low-metallicity galaxies in the IR-submm range,
which are: 
a broadening of the IR peak, which implies a warmer dust component;
an excess of emission in the submm ($\sim$500 $\mu$m), that causes a flattening of the submm/far-IR slope;
and a very low intensity of polycyclic aromatic hydrocarbon emission features. 
}
  {The SEDs of a sample of 27 simulated dwarf galaxies were calculated using the GRASIL-3D radiative transfer code.
This code has the particularity that  
it separately treats the radiative transfer through dust grains within molecular clouds and within the cirrus, the dense and diffuse components of the gas phase, respectively.
The simulated galaxies  have stellar masses ranging from $10^6$ - $10^9$ M$_{\odot}$, and were obtained from a single simulation run within a Local Group environment with initial conditions from the CLUES project.}
  {We report a study of the \textit{IRAS}, \textit{Spitzer}, and \textit{Herschel} bands luminosities,
and of the star formation rates, dust, and gas (HI and H$_2$) mass contents.
  We find a satisfactory agreement with observational data, 
 with GRASIL-3D naturally reproducing  the specific spectral features mentioned above.}
   {We conclude that the GRASIL-3D two-component dust model gives a physical
interpretation of the emission of dwarf galaxies:
 molecular clouds and cirrus represent the warm and cold dust components, respectively, needed to reproduce observational data.}


\titlerunning{IR-submm emission from simulated CLUES dwarf galaxies}
\authorrunning{Santos-Santos et al.}

   \maketitle
%

\section{Introduction}
\label{Intro}

Dwarf galaxies are currently among the less evolved galactic environments: 
they are places where active star formation  is most likely to be found, and
where metallicity is low, and even very low. 
Dwarfs are therefore systems of low stellar mass and low metallicity.
These characteristics have recently deepened the interest in  
 studying  star-forming dwarfs 
as ideal laboratories  to possibly 
anticipate some details of star formation at high redshift.
Metallicity (the mass fraction of heavy elements in the interstellar medium; ISM) is  a key parameter when studying galaxy evolution, and it determines the dust mass fraction in the gas.
Dust forms from the available heavy elements from supernova core-collapse and the outflows from low-mass stars 
\citep[see for example][]{TodiniFerrara:2001,Gomez:2012a,Gomez:2012b,Indebetouw:2014,Rowlands:2014,Matsuura:2015}.

The dust cycle in the ISM is a crucial piece in the process of star formation.
Star-forming regions are embedded in the densest pieces of the ISM, the so-called molecular clouds (MCs).
Dust grains therein are heated by  energetic UV photons that are emitted by young stars \citep{Kennicutt:1998,Kennicutt:2009},
which are particularly susceptible to be absorbed and reemitted in the IR spectral range.
Dust absorption of UV photons in MCs retards molecule dissociation. In this way,
it is an essential piece in the process of ISM cooling, thus facilitating star formation.
Dust in the diffuse gas (cirrus) absorbs light from older stars. It is a matter of debate whether MCs
are also UV photon emitters, or if no UV radiation can escape from them 
\citep[see for example][]{Boquien:2011,Bendo:2012}.

Dust emission from dwarf galaxies  differs considerably from that of more metal-rich galaxies.
They harbor warmer dust, have lower dust-to-gas ratios, present lower polycyclic aromatic hydrocarbon (PAH) band emission, have broader IR spectral energy distribution (SED) peaks, and show a flattening of the far-IR/submm slope
(the so-called submm excess). 
The all-sky {\it IRAS} infrared telescope, 
 which revolutionized our idea of the infrared Universe,
was
  the first to detect warmer dust in galaxies 
\citep[e.g.,][]{Helou:1986,Melisse:1994,Galliano:2003}.
It was followed up by the {\it Spitzer} mission 20 years later, which enabled a
deeper
study of the precise sources found previously by  {\it IRAS}, and it ratified its warm dust findings
\citep[e.g.,][]{Galliano:2005,Rosenberg:2006,Cannon:2006,Galametz:2009}.
More recent investigations have reported broad and flat IR SEDs in dwarf galaxies, further showing that all the peculiarities described previously are more dramatic
 in the most actively star-forming dwarfs
 \citep[e.g.,][]{Boselli:2010, Boselli:2012, Smith:2012a, Ciesla:2014}.
 However, no very low-metallicity ($\rm 12 + log(O/H)\lesssim 8.0$) galaxies were the subjects of these studies.

An important step forward has been made thanks to the {\it Herschel} guaranteed time key
program on dwarf galaxies, the so-called Dwarf Galaxy Survey \citep[DGS,][]{Madden13},
which includes the galaxies with the lowest metallicity in the local Universe.
Subsequent systematic studies \citep{RR13,RR14,RR15} essentially confirmed previous results.
 In addition, the {\it Herschel} Astrophysical Terahertz Large Area Survey \citep[$H$-ATLAS, ][]{Eales10} provided an unbiased view of the dusty Universe.
In particular, a dust-selected sample of local galaxies was made available for the first time, analyzed at high resolution with the sensitivity of {\it Herschel}.
This is the {\it Herschel}-ATLAS Phase-1 Limited-Extent Spatial Survey, 
\citep[HAPLESS][]{Clark:2015},which spans a range of  $\rm 7.4 < M_* < 11.3 \log_{10} M_{\odot}$ in stellar masses,
therefore containing dwarf galaxies. The analysis of this survey revealed the diverse IR properties of local dust-selected galaxies,
including again SED IR peak broadening and slope flattening.

In these pioneering analyses of dwarf galaxies it is generally assumed 
 that these characteristics may be evidence of either an additional warm dust component or a lower-than-Galactic dust emissivity index taking $\beta<2.0$ values, or both.
Specifically, flux densities are modeled as modified black bodies following
\begin{equation}
 F_{\nu} \propto(\frac{\lambda}{\lambda_0})^{-\beta_{\rm}} B_{\nu}(\lambda,T),
\label{ModBB}
\end{equation}
where $\beta$ is the effective slope in the FIR-submm region,
$\lambda_0 \approx 100 \mu$m is the reference wavelength, 
and
 $B_{\nu}(\lambda,T)$  is the Planck function.
In order to fit the data, observers therefore change their modeling in the FIR-submm range  relative to the standard fixed $\beta=2$,  
which is appropriate for normal galaxies
 \citep[e.g.,][]{Boselli2012, RR13}
and/or add new ad hoc additional dust components to it \citep[double black bodies at different temperatures, see e.g.,][]{Galametz:2012, RR14,RR15,Clark:2015}.
In particular, this latter method is needed 
to fit the IR peak broadening that is observed at wavelengths lower than $\lambda \sim 70\,\mu$m. This excess of mid-IR (MIR) emission cannot  be accounted for through a unique modified black body fitting.
For example, HAPLESS analyses find that the flux density modeling through Eq. \ref{ModBB} 
with fixed $\beta=2$ underestimates fluxes at 100 and 500 $\mu$m, 
while overestimating them at 160 $\mu$m. Therefore these analyses use a two-temperature component with fixed $\beta=2$.
On the other hand, \citet{RR15} use a second warmer dust component to cure 
their FIR-submm dust modeling of a lack of photons in the MIR range of some DGS galaxies, when necessary.

These empirical modelings have provided very valuable estimates of 
dust masses, temperatures, and emissivity indices of their analyzed galaxies.
A step further to solve the problem of dwarf galaxy SED peculiarities can be gained
by analyzing dwarf emission in a controlled
scheme, that is to say, when some information
is available about
 the nature and intrisic SEDs of the dust heating sources and on the dust components'
(cirrus + molecular clouds) properties
and space configurations.
This is the case of hydrodynamical simulations.


Two different approaches, both requiring completion with a radiative transfer code to calculate dust effects on the SEDs, can be followed in this regard.
A first procedure is through simulations that include dust formation and evolution in a self-consistent
manner 
\citep[improving one and two-zone models, e.g.,][in that they provide spatial structure]{Dwek98, Lisenfeld98, HirashitaFerrara2002, Calura:2008, Asano:2013}.
These models 
\citep[e.g.,][]{Bekki:2013, Bekki:2015a, Bekki:2015b, McKinnon2016, Aoyama2017} 
include dust formation from stellar evolution and supernova (SN) explosions,   
dust evolution processes  such as grain growth and destruction in the ISM, and dust effects
on galaxy evolution, such as radiation fields effects, star formation, and 
dust-enhanced H$_2$ formation.
The effects of dust in galaxy evolution
 could be important, and studying them self-consistently through hydrodynamical
simulations is a very good approach. This emerging method is still in its infancy, and 
considerable work remains to be done.
Ultimately it will provide a refined estimate of dust properties (e.g., composition,
size distribution, and dust-to-gas mass ratio) in terms of evolutionary paths,
and also in terms of redshift.  At the moment, this method gives different possible solutions that
need to be tuned against observational data.

A second possibility is to study dust effects on the SEDs of simulated galaxies
from the  outputs of standard hydrodynamical simulations that do not include dust formation and evolution,
 but nevertheless produce realistic galaxies
at given redshifts.
In order to study the SEDs of  local (i.e., at redshift $z$=0) dwarf galaxies, 
their dust properties at $z$=0 are needed.
These can be well approximated 
by methods other than using simulations where dust formation and evolution is self-consistently implemented, for example,
following the extensively used models
by \citet{Weingartnera:2001, Draine:2003, Draine:2007} or \citet{Zubko04}, among others,
 which have been thoroughly examined to reproduce observational data.

A few codes exist that interface with the outputs of
hydrodynamical simulations to solve the radiative transfer through
dust and
 predict a multiwavelength SED for simulated
galaxies.
For example,
SKIRT and its updated version \citep{Baes:2002,Baes:2005,Baes:2011,Steinacker:2013},
  SUNRISE
\citep{Jonsson:2004, Jonsson:2006, Jonsson:2009}, 
RADISHE
\citep{Chakrabarti:2008, Chakrabarti:2009}, 
and Art2
\citep{Li:2007, Li:2008, Yajima:2012}
 all use Monte Carlo
techniques to follow the radiation of photons  through  the
 ISM and  calculate  a global radiation field and
hence  the  dust reemission.
In addition,  SUNRISE
includes the treatment of star-forming  regions using the dust
and photoionization code MAPPINGSIII \citep{Groves:2008}.

In a somewhat different context, the dust radiative-transfer code
GRASIL \citep[][hereafter S98 and S99, respectively]{Silva:1998,Silva:1999} has been used highly
successfully for many years,  
in particular in combination with  semianalytical models of galaxy formation such as  GALFORM
 \citep{Cole:2000, Granato:2000, Baugh:2005,
Lacey:2008},  MORGANA  \citep{Monaco:2007,
 Fontanot:2008, Fontanot:2009}, and    ABC \citep{Granato:2004, Silva:2005, Lapi:2006, Cook:2009}.
The 3D version of GRASIL, GRASIL-3D \citep[see][hereafter DT14]{Dominguez:2014},  can 
be interfaced with the outputs of hydrodynamical codes of galaxy formation,
producing observation-like multiwavelength SEDs of simulated galaxies while
inheriting GRASIL code strengths.

Following GRASIL, some GRASIL-3D particular strengths relative
to the codes quoted above can be summarized as follows. 
i) The radiative transfer is not solved
through Monte Carlo methods, but in a grid. 
ii) 
The code separately treats  the radiative transfer in molecular
clouds (MCs) and in the diffuse cirrus component, whose dust compositions are
different. 
We note that MCs are not resolved in simulations, and therefore a subresolution modeling is introduced.
iii) It takes into account the fact that younger stars are associated
with denser ISM environments by means of an age-dependent dust reprocessing of stellar
populations (note that GRASIL has been the first model to do
so), mimicking through the $t_0$
parameter the time young stars are enshrouded within MCs before their destruction. 
iv) It includes a detailed
non-equilibrium calculation  for dust grains with diameter
smaller than $a_{flu} \sim 250$ \AA,
allowing for
a proper treatment of PAH features, which dominate the MIR in
some cases.

The aim of this work is to provide a physically based explanation to the particular
emission patterns dwarf galaxies show in the IR-submm range\footnote{We do not intend  to provide  fittings to observed  individual dwarf galaxy  IR SEDs, since 
the precise baryon 3D distributions and star formation histories of simulated galaxies do not have free parameters.
}.
As described above, these can be summarized as \citep{RR13}
(1) a broadening of the IR peak of the SED, which implies  a warmer dust component;
(2) an excess of emission in the submm ($\sim$500 $\mu$m) that causes a flattening of the submm/FIR slope;
and (3) a very low intensity of PAH emission features.
This is done by studying the IR SEDs of a sample of 27 star-forming dwarf galaxies that are identified in a hydrodynamical simulation of the local Universe. 
This simulation does not include dust evolution self-consistently, but as we show, it produces realistic dwarf galaxies.
The SEDs have been obtained with GRASIL-3D.

The paper is organized as follows:
 Section \ref{method} describes the simulation  and the important characteristics of the  GRASIL-3D radiative transfer code. 
 Sections \ref{sample} and \ref{sec:obs} present the sample of simulated star-forming dwarf galaxies  and the observational galaxies with which the results are compared.
  Sections \ref{sec:prop} and \ref{sec:emi} show the general properties and the IR and submm emission of the sample of simulated galaxies.
 In Sect. \ref{sec:disc} we discuss the results with regard to the physics that drives the existence of an additional warm dust component and the effects of varying 
the free parameters of  
 GRASIL-3D.
 Finally, our study is summarized in Sect. \ref{conclu}.



\section{Method} \label{method}
\subsection{Simulation}
\subsubsection{Initial conditions and run: the CLUES project}
\label{CLUES}

The sample of simulated dwarf galaxies is taken from a single simulation with initial conditions from the Constrained Local UniversE Simulations (CLUES) project\footnote{\tt http://clues-project.org/} 
\citep{gottloeber10, yepes14},
 where peculiar velocities obtained from catalogs are imposed as constraints on the initial conditions in order to simulate a cosmological volume that is representative of our local Universe. In these simulations structures like the Virgo cluster are always reproduced and the Local Group forms in the correct environment.
On  smaller scales,  the distribution of structure is essentially random.

Several dark matter-only realizations are run until a  Local Group analog (a Milky Way-M31  binary group, hereafter LG) is found. Then the zoom-in technique is applied, which means that the LG region is resimulated with baryons and at a higher resolution. The resimulation includes 4096$^3$ effective particles in a sphere with radius $\sim 2h^{-1}$Mpc around the LG. The mass resolution of particles is m$_{\rm star}$=1.3$\times$10$^4$M$_{\odot}$, m$_{\rm gas}$=1.8$\times$10$^4$M$_{\odot}$, and m$_{\rm dm}$=2.9$\times$10$^5$M$_{\odot}$, and the gravitational softening lengths are $\epsilon_{\rm bar}$=223pc between baryons and $\epsilon_{\rm dm}$=486pc between dark matter particles.
The cosmology used is $\Lambda$CDM with WMAP3 parameters, that is, $H_0=73$ km s$^{-1}$ Mpc$^{-1}$, $\Omega_{\rm m}=0.24$, $\Omega_{\Lambda}=0.76$, $\Omega_{\rm baryon}=0.04$, and $\sigma_8=0.76$.

\subsubsection{GASOLINE}\label{gasoline}
This CLUES simulation has been evolved using the parallel N-body+SPH tree-code \verb,GASOLINE, \citep{wadsley04}, which includes gas hydrodynamics and cooling, star formation, energy feedback, and metal enrichment to model structure formation. 
We briefly describe here  the most important implementations of this well-tested code (for details see \citealt{stinson06} and \citealt{governato10}).

The simulation 
follows  the physics used in \citet{governato10} and \citet{guedes11}. 
Feedback recipes are tuned  to match the stellar-to-halo mass relation at one stellar mass. 
When gas becomes cold and dense, stars are formed according to a Schmidt law with a star formation rate $\propto\rho^{1.5}$. When these stars die,  they release energy and metals into the surrounding ISM.
The total mass and number of stars that explode as SNe
 is calculated using stellar lifetime calculations from \citet{Raiteri96}.
 It is assumed that
only stars between 8 and 40 M$_\odot$ explode as SNII, 
and the mass of a binary system that can eventually explode as SNIa is between 3-16 M$_\odot$.
Stars more
massive than 40 M$_\odot$ are assumed to either collapse into black 
holes or explode as SNIb. Regardless of this, only few stars
form with masses greater than 40 M$_\odot$,  thus the impact on the
feedback is minimal.
 The energy feedback by SNII is implemented by means of the blastwave formalism \citep{stinson06}, where
a fixed fraction of the canonical $10^{51}$ erg per SN explosion is released to the ISM.  
The metals produced in these stars are released as the main-sequence progenitors die,
 and the metals are distributed to the gas within the
blast radius. Iron and oxygen
are produced in SNII according to the analytic fits used in \citet{Raiteri96}
 using the yields from \citet{Woosley95}:
  M$_{\rm Fe}$ =2.802 $\times$  10$^{-4}$M$^{1.864}_*$ and 
M$_{\rm O}$ =4.586 $\times$ 10$^{-4}$M$^{2.721}_*$.

Feedback from SNIa also follows the \citet{Raiteri96} model. Each SNIa produces 0.63 M$_\odot$ iron and
0.13 M$_\odot$ oxygen \citep{Thielemann86} and the
metals are ejected into the nearest gas particle. 
Furthermore, the significant feedback contribution of stellar winds is also taken into account.
Stars with masses of between 1 and 8
M$_\odot$ return a fraction of their initial mass,
which is determined
using the function derived by \citet{Weidemann87}. The returned gas
has the same metallicity as the star particle. Metal diffusion between gas particles is implemented as described in \citet{shen10}. 
Finally, \verb,GASOLINE, also accounts for the effect of a uniform background radiation field on the ionization and excitation state of the gas.

 The formation and evolution of dwarfs in the LG neighborhood has been analyzed in zoomed simulations that are based on the same initial conditions as used in our paper, but performed with the GADGET code \citep{BenitezLlambay13, BenitezLlambay15, BenitezLlambay16}.
This particular CLUES simulation has been studied in \citet{SantosSantos16}, where it was shown that its intermediate-  to high-mass disk-like galaxies have a correct mass distribution and angular momentum content.

\subsection{GRASIL-3D}
\label{GRASIL3D}
 
To develop the GRASIL-3D radiative-transfer code, DT14 have followed the main
characteristics and scheme of GRASIL. We briefly recall the main features
here. For a full description we refer to DT14, S98, and  S99.

 To solve the radiative transfer, 
GRASIL-3D uses a Cartesian grid whose cell size is set by the smoothing length
used in the simulation code. 
The gas is subdivided into a dense phase (fraction $f_{mc}$
of the total mass of gas) associated with young stars
(star-forming molecular clouds, MCs) and a diffuse phase 
(cirrus) where more evolved (free) stars and MCs
are placed.
Young stars leave their parent clouds on a timescale $t_0$.
The MCs are represented as spherical clouds with 
a certain
optical depth
 containing an inner stellar source of radiation, whose
radiative transfer through the MC is calculated following \citet{Granato:1994}. The radiative
transfer of the radiation emerging from MCs and  from free
stars is then computed through the cirrus dust.

\subsubsection{Gas fractions in MCs and cirrus}

The fraction of gas in MCs, $f_{mc}$, is 
calculated locally  by implementing a
subresolution model based on the assumption that MCs are defined by a density threshold $\rho_{mc, thres}$ 
\citep[e.g.,][]{Elmegreen:2002,Hennebelle:2012}, and using a theoretical log-normal probability distribution function (PDF) for the gas
densities.
The PDF  is characterized by a density parameter $\rho_0$ and a dispersion $\sigma$.
They define the PDF volume-averaged density as
$ <\rho>_V \equiv\rho_0 e^{\sigma^2/2}$,
 which we in turn identify with the SPH density of the \textit{i-th} gas particle (simulation output),
$ <\rho>_V = \rho_{\rm gas}(\vec{r_i}) $.
In this way,
$\rho_0$ can be calculated for each gas particle,
leaving only two parameters to control 
the calculation of $f_{mc}$: the gas density threshold $\rho_{mc, thres}$, and the PDF dispersion $\sigma$. 
The diffuse gas fraction of a given particle is then $(1 - f_{mc})$.
This PDF approach is 
suggested by small-scale ($\sim$ 1 kpc) simulations 
\citep[see, for example,][]{WadaNorman:2007,Federrath:2008,Brunt:2009,Hennebelle:2012}
and is supported by recent  high-resolution {\it Herschel}  observations of nearby MCs \citep[e.g.,][]{Schneider12}.

 Values for  $\rho_{mc, thres}$ vary in the literature.
    For example,
    $\rho_{mc, thres}$ = 100 H nuclei cm$^{-3} \approx$ 3.3 M$_{\odot}$ pc$^{-3}$ in \citet{Tasker:2009},
    and $= 35$ H nuclei cm$^{-3} \approx 1$ M$_{\odot}$ pc$^{-3}$ 
    in \citet{BallesterosP:1999,BallesterosP_Scalo:1999}. 
    Values for $\sigma$ have been calculated to
    range from 2.36 to 3.012 in \citet{Wada:2007},
 while \citet{Tasker:2009} give $\sigma$ = 2.0.     
 In DT14 the  range  for the density threshold 
     is  taken to be $\rho_{mc,thres}$ = 10 - 100 H nuclei cm$^{-3} \approx 3.3 \times 10^8 - 3.3 \times 10^9  $M$_{\odot}$ kpc$^{-3}$, while the PDF dispersion takes either 2.0 or 3.0. 
     These are the values used here.
A useful check for the  choice of these parameters can
be made by comparing  
the final calculated average HI and H$_2$ contents of the galaxy or sample of galaxies, with observations.

\subsubsection{Stellar emission}

The luminosity of the young stellar
populations placed inside each active MC
 is modeled in GRASIL 
by linearly decreasing the fraction $f$ of the stellar populations' energy radiated inside the cloud with its age $t$:
\begin{equation}
f(t) = \left\{ \begin{array}{ll}
 1         & \textrm{  $t \le t_0$} \\
 2 -t/t_0   & \textrm{ $t_0 < t \le 2 t_0$ } \\
 0         & \textrm{$t > 2 t_0$}
               \end{array} \right.
\label{fracyoung}
\end{equation}
Here $t_0$ is a free parameter representing the time taken for stars to escape the MCs where they were born, 
and it mimics MC destruction by young stars. 
The same approach is adopted in GRASIL-3D.
In this way,
only stars with ages<2$\times t_0$ are eligible to form part of the MC heating engine, with probability $f(t)$.
Observationally, $t_0$ takes values on the order of the lifetime of the most massive stars, from 3 to 100 Myr, and it varies with the density of the surrounding ISM, being higher in the densest environments. 
Typical values were found by
comparison  to local observations in S98, yielding $\sim$
2.5 to 8 Myr for normal spiral galaxies, and between 18
and 50 Myr for starburst galaxies.
The contributions to the
heating of cirrus dust come from free stellar particles with ages
$t > 2 t_0$, as well as from stars with $t > t_0$ supplying  $1- f(t)$ of
their radiation.

\subsubsection{Dust-to-gas mass ratio}
\label{DGratio}
The dust content in MCs and in the cirrus is computed from a dust-to-gas mass ratio $\delta$.
In this work, $\delta$ has been set to vary with 
 gas  metallicity Z$_{gas}$ 
following the broken power-law in terms of oxygen abundances proposed in Table 1 of \citet{RR14}.
This approach
is needed 
to fit the 
very low
dust masses  observed in low-metallicity (12+log(O/H)$\lesssim$8.0) environments or galaxies
\citep{Lisenfeld98, Galliano:2005, Hunt2005, Galliano:2008, Engelbracht2008, Galametz2011, Galliano:2011,  Izotov:2014},
and it is consistent with dust and PAH evolution models \citep[e.g.,][]{Hirashita:2002, Calura:2008, GallianoDwek:2008, Asano:2013, Seok:2014}.
Since in GRASIL-3D each particle has
 a different gas mass and metallicity, the dust-to-gas mass ratio is a local quantity that varies from cell to cell.
Adopting 12+log(O/H)$_\odot$ = 8.69 \citep[i.e. Z$_\odot$ =0.014,][]{Asplund09} solar values,
 we obtain
\begin{eqnarray}
\delta(Z_{gas, k}) = \frac{D}{G} =  \frac{1}{10^{a_1}}\left( \frac{Z_{gas, k}}{Z_\odot}\right)^{\alpha_1} \quad \rm for\ Z>Z_t \\
\delta(Z_{gas, k}) = \frac{D}{G} =  \frac{1}{10^{a_2}}\left( \frac{Z_{gas, k}}{Z_\odot}\right)^{\alpha_2} \quad \rm for\ Z\leq Z_t \nonumber
\label{dustfrac}
\end{eqnarray}
where $\alpha_1$=1.0, $a_1$=2.21, $\alpha_2$=3.0, $a_2$=0.77, 
and $Z_{gas, k}$ is the  mass fraction of gas metals at the $k$-th grid cell.
The break in the metallicity dependence occurs at Z$\rm_t$=0.0026 (i.e., 12+log(O/H)=7.96),
and the result is given in solar units, where $\delta_\odot$=D/G$_\odot$=1/162 according to \citet{Zubko04} for a dust composition like ours (see next subsection).
We adopt a slope of 1 at high metallicities ($\alpha_1=1$), which has been shown in many studies to adequately represent this regime
\citep{James2002, Draine:2007, Galliano:2008, Leroy:2011},
and  a slope of $\alpha_2=3.0$ at low metallicities 
(we do not use exactly the same value as that given in \citet{RR14} but a simpler value that is inside their errors).
This broken power-law scaling  of $\delta$ with metallicity gives satisfactory final dust masses compared to observational estimates, as we show in Sect. \ref{dgratio},
 while using a linear scaling ($\alpha_1 = \alpha_2 = 1$) at any metallicity does not work so well.
However, we note that
the problem of the D/G vs Z assignation at low metallicity is still not solved because of 
 the uncertainties in determining  both the dust and H$_2$ masses of observed very low-metallicity galaxies.

The dust-to-gas mass ratio is also relevant for calculating the radiation transfer through MCs since it determines the optical depth of the cloud:
\begin{equation}
\tau_{mc} \propto \delta(Z_{gas, k}) \frac{m_{mc}}{r_{mc}^2},
\label{taumc}
\end{equation}
where $m_{mc}$ and $r_{mc}$ are the mass and radius of individual MCs, respectively, both free parameters  in GRASIL-3D.
The proportionality constant depends on the dust model.
Since both $\tau$  and the luminosity from the central source can vary in each grid cell, the  radiative transfer is calculated separately for each of them even if $t_0$ is taken to be constant.

\subsubsection{Dust model}\label{dustmodel}
The dust in GRASIL-3D is assumed to consist of a mixture of 
different grain types:
carbonaceous and silicate spherical grains, 
where graphite grains smaller than $a_{flu}=$250$\AA$ are considered as PAHs. 
These in turn consist of a mixture of neutral and ionized particles.
This simple composition has been shown to appropriately reproduce observational constraints coming from the average interstellar extinction, thermal IR emission, and interstellar abundances in the Galaxy, LMC and SMC \citep{Weingartnera:2001,Zubko04}.
A different size distribution is adopted for each grain type, following the functional form given in \citet{Weingartnera:2001}, and updated in \citet{Draine:2007}.
The optical properties come from \citet{Laor:1993} and \citet{Li:2001}.

A different dust composition is assumed in the diffuse ISM and in MCs.
The difference resides in the 
PAH abundance or index $q_{\rm PAH}$, defined as
the percentage of dust mass contributed by PAHs containing less than $10^3$ C atoms.
Previous studies have discretized this index into seven values
\citep[see][]{Weingartnera:2001, Draineb:2007, Draine:2007},
 each characterizing a specific dust model, where the highest value (4.58\%) matches the Galactic dust.
Here, we use a low value of $q_{\rm PAH}$ = 1.12\% for the cirrus dust \citep[as in model MW3.1\_10 from][]{Draine:2007, Draineb:2007},
 as suggested by observations of low-metallicity galaxies that show very low PAH band emission
  \citep{Engelbracht2005, Ohalloran2006, Draineb:2007, Smith2007b, Gordon2008, Rosenberg:2008, Hunt2010, WuR2011, Ciesla:2014}
and by PAH evolution models \citep{Seok:2014, Bekki:2013}.
 This has been implemented globally for all grid cells, and we consider it a suitable approximation since
the use of one value for $q_{\rm PAH}$ or another only  affects the 
intensity of the particular mid-infrared PAH emission features, and not 
the far-infrared luminosity.
In MCs however, the PAH abundance is decreased 1000 times relative to the diffuse component, as tuned by \citet{Vega:2005} to fit MIR properties of local actively star-forming galaxies.
The differences between using this low $q_{\rm PAH}$ cirrus value or the standard Galactic value are discussed in Sect. \ref{ParVar}.

The temperature distribution is defined for each grain type (silicate, graphite, or PAH) by its particular size distribution. In the case of grains with sizes $a> a_{flu}$, each interval in grain size corresponds to  emission from a 
 modified black body of a given temperature $T$, with an emissivity index of $\beta$=2 defining its FIR/submm slope.
For PAH grains, in contrast,  no unique temperature is defined.
PAHs of a given size are not able to reach temperature equilibrium due to stochastic heating, and span instead 
 a  wide distribution of temperatures \citep{Guhathakurta:1989, Siebenmorgen:92}. 
 The final emission of the MC and cirrus dust components in a given grid cell is the sum of all the temperature contributions coming from each grain size interval corresponding to each grain type.

Throughout the paper we  use the respective maxima location of the 
MC and cirrus dust components
as an estimate of their effective global temperatures. 
 However, it is worth noting that
  the typically used  `cold' or `warm' terms for dust emission 
do not strictly apply for GRASIL-3D SEDs 
(nor for real galaxies)
as they represent
  a very simplified description of a more complex situation.

\subsubsection{Choice of GRASIL-3D free parameter values}

According to the previous subsections,
 the free parameters in GRASIL-3D are
i) the escape timescale from MCs, $t_0$; ii) the threshold density for MCs, $\rho_{mc, thres}$;
iii) the log-normal PDF dispersion parameter, $\sigma$; and iv) the ${m_{mc}}/{r_{mc}^2}$
combination, determining the optical depth of individual MCs.
It is worthwhile to recall that the simulation provides the geometry of each galaxy, the star formation history (SFH), metal enrichment history, and stellar and gas fractions, in such a way that no further parameters are needed to describe them.

The following set of parameter values have been used and 
considered as of reference in this study:
 $t_0$=40 Myr, 
$\rho_{mc,thres}=3.3 \times 10^{9}$ M$_\odot$kpc$^{-3}$, $\sigma$=2., $m_{mc}$=10$^6$ M$_\odot$, and $r_{mc}$=14 pc.

The choice of a high $t_0$ value comes from following the results of S98 for star-bursting galaxies,
which seems a good approximation since as we show below, our simulated dwarfs are particularly picked according to their recent star formation.
The PDF parameters chosen provide the best match to observational data of H$_2$ molecular gas masses of nearby low-mass galaxies when using a metallicity-dependent CO-to-H$_2$ conversion factor (see Sects. \ref{SFRH-H2} and \ref{HIH2mass}). Finally, the mass and radii of single molecular clouds are set to the values that best match observed SEDs of local galaxies in S98. These numbers are within the  observational estimates from MCs in our Galaxy.
In Sect. \ref{ParVar} we discuss the effects of parameter variation on the resulting SEDs in detail.

GRASIL-3D has a general applicability to the outputs of
simulated galaxies, and it has been used in a variety of contexts 
to analyze their SEDs.  For example,
DT14 have analyzed normal spiral galaxies and compared the resulting SEDs to different observational surveys.
\citet{Obreja:2014} have tested star formation rate tracers used by observers against mass-weighted
quantities, obtaining a satisfactory agreement.
\citet{Granato:2015} have analyzed the early phases of galaxy cluster formation in hydrodynamical simulations,
comparing their  IR emission to observations.
\citet{Buck2016} have studied the nature of clumps in simulated massive disk galaxies.
\citet{Goz2016} have evaluated the possible correlation of IR luminosity in \textit{Spitzer} and \textit{Herschel} bands with several galaxy properties.

%
 \begin{figure}
          \resizebox{\hsize}{!}{\includegraphics[scale=0.6]{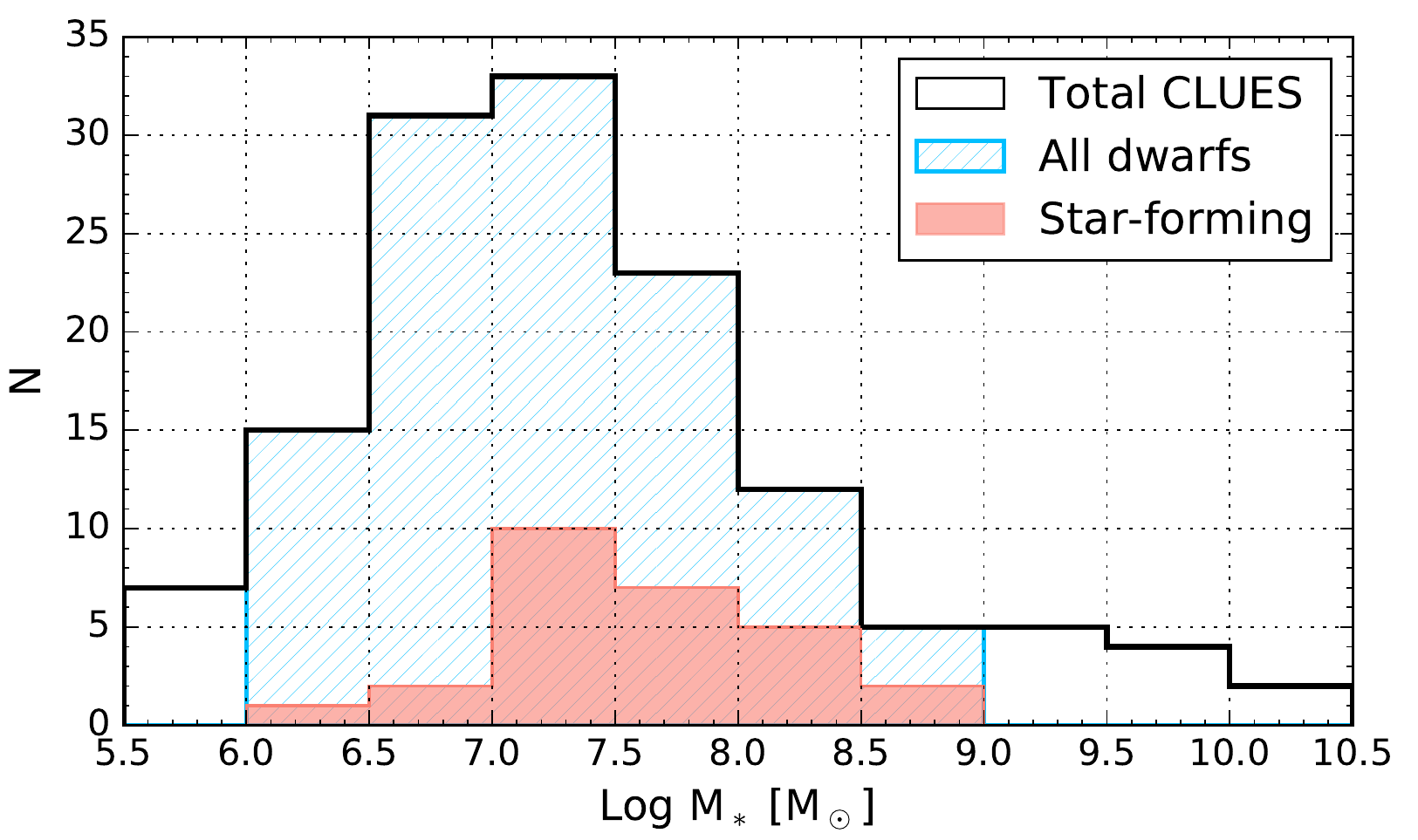}}
   \caption{Stellar mass distributions of the CLUES total (black outline), dwarf (blue pattern), and star-forming dwarf (red) galaxy samples. }
              \label{FigMass}%
 \end{figure}

 \begin{figure}
     {\includegraphics[width=\linewidth]{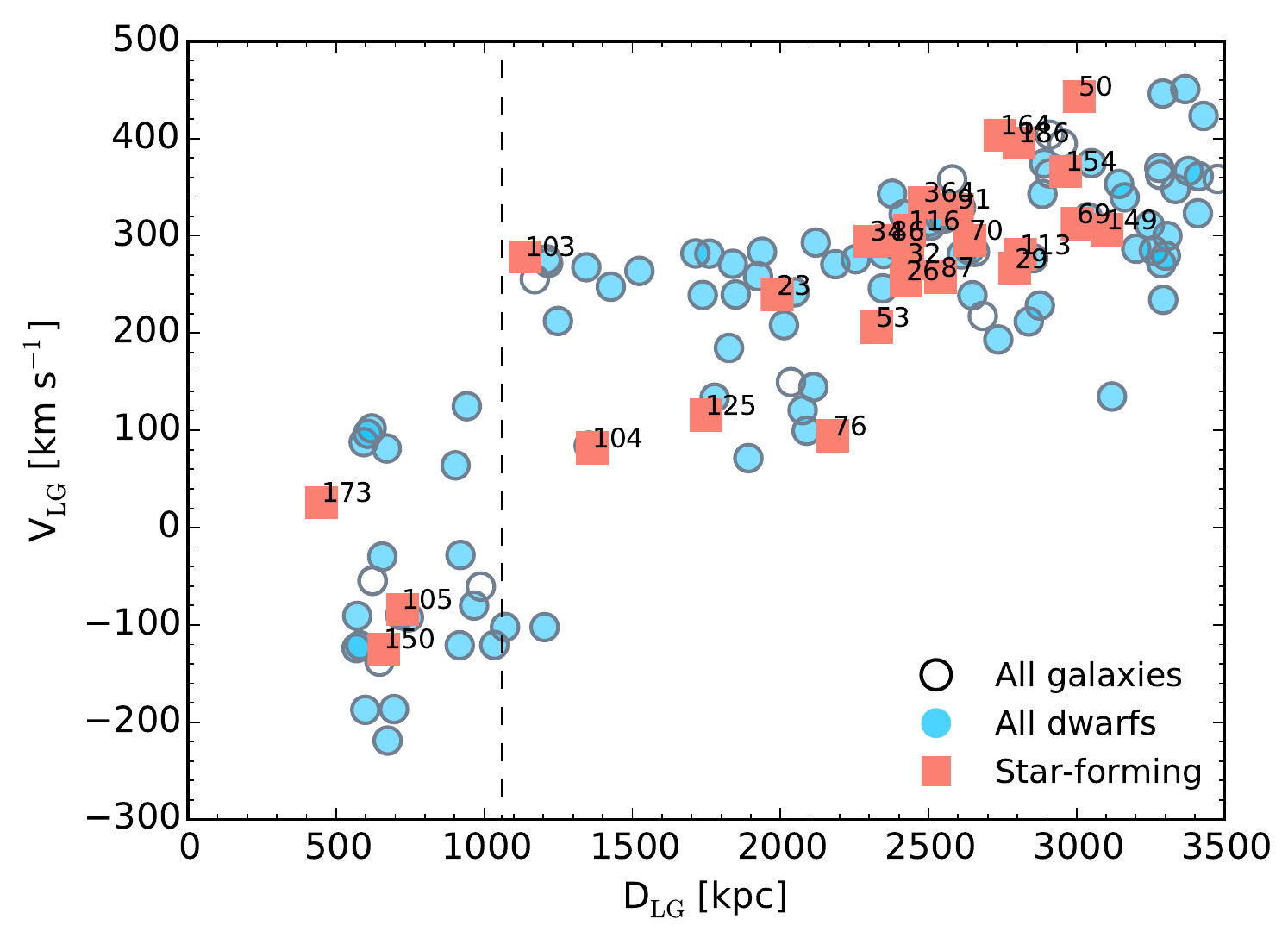}}
   \caption{Distance to the CLUES Local Group barycenter versus the velocity relative to this barycenter. The color-coding is the same as in Fig. \ref{FigMass}. The vertical dashed line indicates the observational zero-velocity surface of the
Local Group (R$_{\rm LG}$ = 1060 $\pm$ 70 kpc) as derived in \citet{McConnachie12}.}
              \label{FigDvsV}%
 \end{figure}


\section{Sample of simulated star-forming dwarf galaxies} \label{sample}

The  galaxies from the CLUES simulation were identified with  the Amiga Halo Finder (\texttt{AHF}; \citealt{knollmann09}), where halo masses are defined as the mass inside a sphere containing $\Delta_{\rm vir}\simeq$350 times the cosmic background matter density. 
This search has produced 139 highly resolved dark matter halos, 120 of which host
bound stellar mass objects ranging 10$^6-$10$^9$M$_\odot$, to which we asign the label "dwarf". These may or may not have gas particles, and can be either isolated or satellites.

In this work we are interested in the IR-submm emission of dwarf galaxies.
As is well known,
dust grains absorb the ultraviolet light emitted by very young stars and reemit it at longer wavelengths. 
We therefore focus on those dwarfs that have recent star formation; in particular, dwarfs that contain stellar particles (stellar populations) with ages lower than 80 Myr 
 (this is twice the $t_0$ value we chose). 
This leaves a final sample of 27 dwarf galaxies, all of which contain gas. Their general properties are given in Table \ref{Table1}.

The mass distributions of the CLUES total, dwarf, and star-forming dwarf galaxy samples are given in Fig. \ref{FigMass}. 
 We note that 
the fraction of galaxies in the LG with recent star formation is consistent with observations \citep{Mateo:1998,Tolstoy:2009, Dolphin:2005}. 
In Fig. \ref{FigDvsV} we plot the velocity of each galaxy relative to the center of mass of the simulated LG, that is, the barycenter of the Milky Way-M31 pair, versus their distance to this barycenter. A vertical dashed line indicates the zero-velocity surface of the LG (R$_{\rm LG}$ = 1060 $\pm$ 70 kpc) as derived in \citet{McConnachie12}. 
Galaxies are labeled with their ID number (column 1, Table \ref{Table1}) to allow following them  in all figures throughout the paper.
When compared to real data \citep[see Fig. 5]{McConnachie12},
this plot gives us a first 
satisfactory 
insight into the simulated LG structure, and in particular, into the galaxy
distribution. Considering the local volume to have a radial extent of $\sim$3 Mpc, we find galaxies that are true LG members, galaxies that are close  but not yet bounded, and galaxies that are outer nearby neighbors, rising in velocity with distance as expected due to the Hubble flow. 
The total number of simulated LG members is comparable to the observed number.

The galaxies of this sample present differences in their SFHs, gas masses, sizes, and morphologies, the latter being either irregular, spheroidal, or disky. 
Furthermore, the majority of the local star-forming dwarfs we studied are isolated,
except for
 three (numbers 86, 104, and 150) that are satellites of other more massive galaxies.
Some of the general properties of the sample, listed in Table \ref{Table1}, are studied in Sect. \ref{sec:prop}.


\section{Observational samples to compare our results with}\label{sec:obs}

We report on two types of comparisons with observational data. First, we examine whether the simulated dwarfs have global properties that agree with observational samples, for which we focus  on the SFHs, HI and H$_2$ gas contents, dust masses, and metallicities because these properties have the strongest influence on the resulting SEDs. Second, we compare the IR-submm emission of our simulated sample with observed galaxies that span the same stellar mass and metallicity ranges. This second comparison is the core of this paper, and it is made possible by the recent \textit{Herschel} observations mentioned in Sect. \ref{Intro}.

\subsection{ HI and H$_2$ gas content, star-formation rate, and gas metallicity}   
\label{SFRH-H2}
We compare the HI and H$_2$ masses of the simulated dwarf galaxies to data from the observational studies described below.
For all of these galaxy samples
the H$_2$ content or molecular gas masses are calculated by assuming a constant CO-to-H$_2$ 
conversion factor with
the Galactic value: X$_{\rm CO;MW} =2.0 \times 10^{20}$ cm$^{-2}$ (K km s$^{-1}$)$^{-1}$.
However, 
we note that it has long been argued
that CO may not trace the  molecular gas in dwarf galaxies: the conversion factor may be metallicity dependent 
\citep[see e.g.,][]{Taylor98, Wilson:1995,Boselli:2002, Hunt2005, Leroy:2011,Schruba:2012, Bolatto13}, 
and the use of the standard Galactic value could understimate the molecular gas fraction in low-metallicity galaxies. 
Some of the following samples also provide the H$_2$ masses obtained using a conversion factor of this type.
We  therefore use these samples 
in Sect. \ref{HIH2mass}
to study the 
different GRASIL-3D parametrizations needed to match the molecular gas masses obtained
using a constant and a metallicity-dependent conversion factor.\\

\paragraph{The Dwarf Galaxy Survey  \citep[DGS,][]{Madden13}.}
DGS is a sample of star-forming dwarf galaxies with metallicities ranging from $Z \sim 0.03$
to $0.55\,Z_{\odot}$ and stellar masses covering  $ 3 \times 10^6$ to $3 \times 10^{10}$ M$_{\odot}$
\citep{Madden14}. HI and H$_2$ masses are given in \citet{RR14}, where they compiled data from \citet{Madden13}.
HI masses were corrected to account for the aperture size used to obtain the dust SED. 
 \citet{RR14} also provide the H$_2$ masses of the DGS galaxies obtained using a metallicity-dependent conversion factor of the form $ X_{\rm CO,Z}\propto Z^{-2}$, following \citet{Schruba:2012}.    
Dust masses for DGS are given in \citet{RR15}, where they represent the best fit to data following the dust SED model from \citet{Galliano:2011}.

\paragraph{The Key Insights on Nearby Galaxies: a Far-Infrared Survey with \textit{Herschel} \citep[KINGFISH,][]{Kennicutt11}.}
This sample consists of 61 galaxies, most of them spirals with some early-type and dwarf galaxies.
Their stellar mass range is $\rm 1.7 \times 10^7-4.8 \times 10^{11} M_{\odot}$
 peaking at $\rm M_* \sim 5.0 \times 10^{10} M_{\odot}$, and metallicities reach up to Z $\sim$ 1.20 Z$_{\odot}$.
No particular dust emission features have been found in this sample \citep{RR15}. We therefore use it as a reference with which to compare the specifities of dust emission in lower mass galaxies. HI and H$_2$ masses of KINGFISH galaxies are also given in \citet{RR14}, where they compiled data from \citet{Draineb:2007}. 
 \citet{RR14} additionally provide the H$_2$ masses obtained using a metallicity-dependent conversion factor of the form $ X_{\rm CO,Z}\propto Z^{-2}$.
Dust masses for KINGFISH are estimated  in \citet{RR15}.

\paragraph{The \textit{Herschel}-ATLAS Phase-1 Limited-Extent Spatial Survey \citep[HAPLESS,][]{Clark:2015}.}
This is the first 250 $\mu$m blind survey selected from the \textit{Herschel} Astrophysical Terahertz Large Area Survey \citep[H-ATLAS,][]{Eales10}
on the basis of their (higher) dust mass content. It consists of 42 nearby galaxies, spanning a range in stellar mass from $ 2.5 \times 10^7$ to $2 \times 10^{11}$ M$_{\odot}$
with a peak at M$_* \simeq  3 \times 10^8$ M$_{\odot} $. They are the most actively star-forming galaxies in H-ATLAS,
with $ -11.8 <$ sSFR $< -8.9 \log_{10}$ yr$^{-1}$, and seem to be in an early stage of converting their gas into stars. HI masses are calculated from the  highest-resolution 21 cm
observations available for each galaxy, coming either from the HI Parkes All Sky
Survey  \citep[HIPASS,][]{Meyer:2004}, the Arecibo Legacy
Fast ALFA Survey \citep[ALFALFA, ][]{Haynes:2011}, or from other published literature sources.

\paragraph{\citealt{Leroy08}.}
These authors compiled the available data for 23 nearby star-forming galaxies to study star formation rates (SFRs) and efficiencies. Half of them are HI-dominated low-mass ($\rm <10^{10}\,M_\odot$) galaxies and the other half are large spirals. The HI and H$_2$ masses are calculated from integrating the atomic (THINGS, \citealt{Walter08}) and molecular (HERACLES, \citealt{Leroy09} or BIMA SONG, \citealt{Helfer03}) hydrogen surface densities within 1.5 r$_{25}$.

\paragraph{\citealt{Groves15}.}
The correlation between 
the total gas mass and the IR luminosity is studied for  a 
sample of 36 nearby (D$\sim$10 Mpc) galaxies observed
with the KINGFISH, THINGS, and HERACLES surveys.
These galaxies range from 10$^{6.5}$ to 10$^{10.5}$ M$_\odot$ in stellar mass  and have disk-like
morphologies. 
Molecular gas masses are calculated using HERACLES data by assuming a fixed-to-Galactic-value CO-to-H$_2$ conversion factor and a fixed line ratio of CO(2-1) to CO(1-0) of R$_{21}$ = 0.8 .

\paragraph{APEX Low-redshift Legacy Survey for MOlecular Gas \citep[ALLSMOG,][]{Bothwell:2014}.}
ALLSMOG is designed to observe the CO(2-1) emission line with the APEX
telescope in a sample of local galaxies (0.01 $< z <$ 0.03). These galaxies are from the SDSS data release 7, and their stellar masses are in the
  10$^{8.5} - 10^{10}$ M$_\odot$
range.
HI masses are taken
from either HIPASS, ALFALFA, 
 or from the large collection of HI observations
assembled by \citet{Springob:2005}. 
\citet{Bothwell:2014} also provide for this sample the results of the H$_2$ masses calculated using the metallicity-dependent CO-to-H$_2$ conversion factor of \citet{Wolfire10}.

\paragraph{The Faint Irregular Galaxies GMRT Survey \citep[FIGGS,][]{Begum08}.}
This is a HI imaging survey of  65 extremely
faint nearby dwarf irregular galaxies observed with the Giant Meterwave Radio Telescope. It is a subsample of the \citet{Karachentsev04} catalog of galaxies within a distance of 10 Mpc.
HI masses were derived from global HI profiles. 
\\

We note that 
several galaxies are in common in these samples,
 and  the atomic and molecular hydrogen masses each reference gives can be  different for a same object. We kept all values in our analysis to account for the scatter coming from the discrepancy in these mass determinations.
\\

The SFRs and metallicities of the CLUES star-forming dwarfs were compared to those of galaxies from DGS and KINGFISH.
For these two surveys, SFRs were estimated considering two of the most widely used SFR tracers, the far-ultraviolet (FUV) and the H$\alpha$ luminosities, corrected for
dust attenuation, using either total-infrared (TIR) or 24 $\mu$m luminosities \citep{Kennicutt09, Calzetti10, Hao11, Kennicutt12}. We refer to Section 2.6.1 in \citet{RR15} for details of the estimations.
The metallicities in both samples have been determined in
\citet{Madden13} and \citet{Kennicutt11} 
using the calibration from \citet{Pilyugin05}.

\subsection{IR-submm emission}

We compare our IR-submm emission results to data from DGS, KINGFISH, and HAPLESS.
All three galaxy samples have been observed with \textit{Herschel} in the 70, 100, and 160 $\mu$m PACS
and 250, 350 and 500 $\mu$m SPIRE bands \citep[see][respectively]{RR15,Clark:2015,Dale12}.
Mid-infrared data from IRAC 8$\mu$m and MIPS 24$\mu$m \textit{Spitzer} bands for DGS and KINGFISH \citep[see][respectively]{RR15, Dale:2005},
are also studied. 
We highlight that GRASIL-3D allows the precise calculation of the  luminosities of the simulated galaxies as seen through a very wide range of telescopic filters (with their particular transmission and calibration), allowing for an accurate comparison with the above observational data.


\section{Results: General properties of the sample}\label{sec:prop}

We briefly present here the general properties of 
 the CLUES star-forming dwarf galaxy sample,
  namely, their SFHs, neutral hydrogen mass content, and metallicities, to demonstrate that they constitute a realistic dwarf galaxy sample with properties comparable to observed properties.
  The molecular H$_2$ gas masses as calculated by GRASIL-3D are also compared to observational data to justify the choice of $\rho_{mc,thres}$ made.



\begin{figure}
       \resizebox{\hsize}{!}{\includegraphics{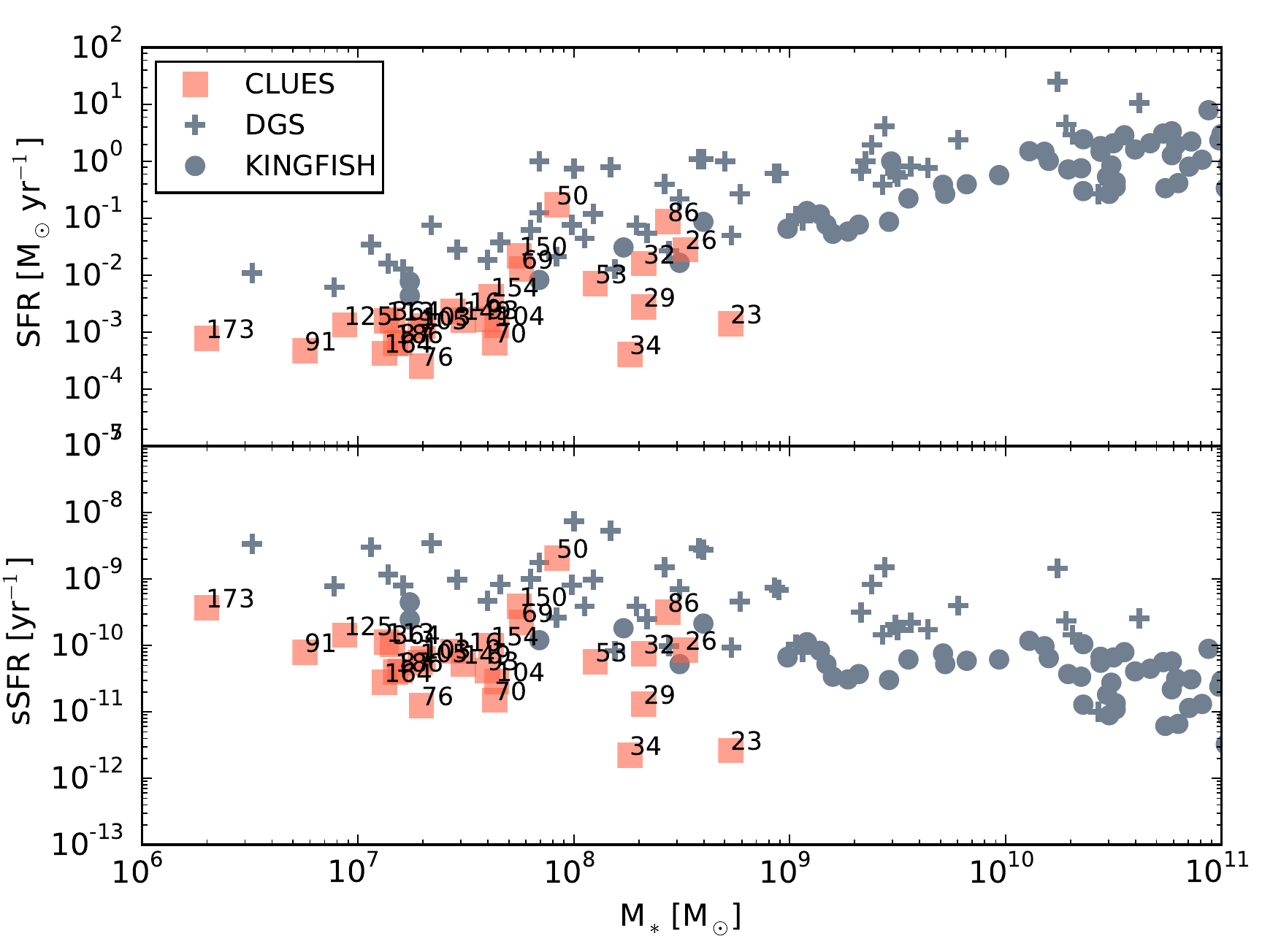}}
   \caption{Star formation rate (upper panel) and specific star formation rate (lower panel) of the CLUES star-forming dwarf galaxies compared to observational data from the DGS and KINGFISH galaxy samples.}
     \label{Figsfr}
    \end{figure}

\begin{figure}
      \resizebox{\hsize}{!}{\includegraphics{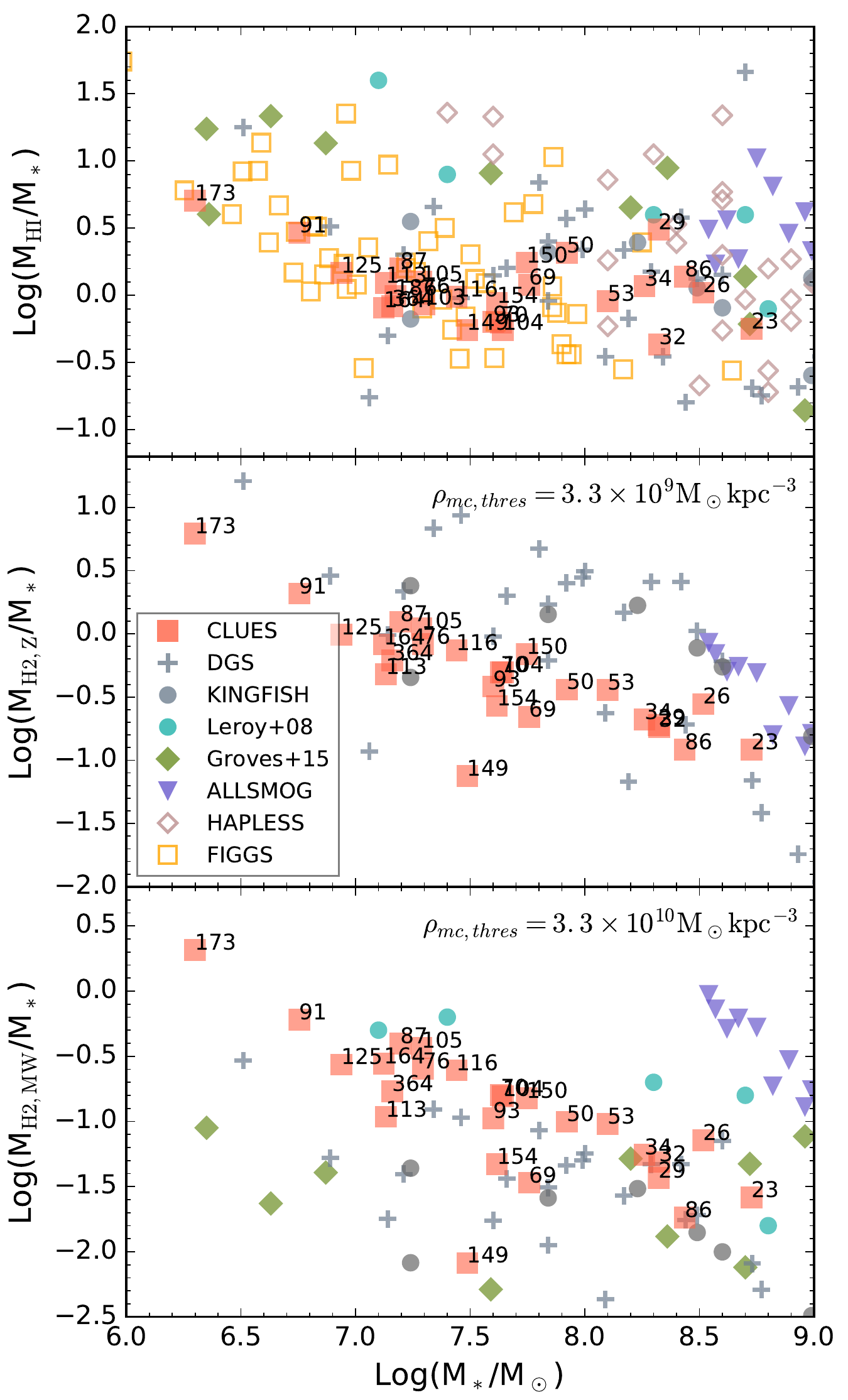}}
   \caption{  M$_{\rm HI}$/M$_*$ (upper panel) and M$_{\rm H2}$/M$_*$ (middle and lower panels) ratios versus stellar mass. CLUES simulated galaxies are shown as red squares, while the various observational samples take the symbols shown in the legend.  HI and H$_2$ masses are calculated by GRASIL-3D based on a log-normal PDF for gas densities.
The middle and lower panels differ in the assumption for the observed galaxies of a metallicity-dependent or constant CO-to-H$_2$ conversion factor, respectively, which in GRASIL-3D means the use of different density threshold values $\rho_{mc,thres}$ in order to match observational data. }
     \label{FigHcontent}%
    \end{figure}



\subsection{Star formation rate history}

The 
SFRs at $z=0$ 
of the CLUES dwarfs were calculated using
the mass  of stars born in the past 100 Myr, a time interval that is consistent with  SFR tracers such as  FUV and 
TIR luminosities, used 
by the observational samples we compare our results to.
Their star formation rate histories (SFHs) 
are depicted in Fig. \ref{FigSFH}
as histograms where each bin represents 100 Myr.
They show
significant diversity, some of them having formed many stars recently, while others have not. 
 In general, all the simulated dwarf galaxies present a bursty SFH, with an average of $\sim$2.5 important bursty episodes per galaxy, which are generally short lived ($\sim$200 Myr).
 
In Fig. \ref{Figsfr} we show the SFR and specific SFR (sSFR) at $z=0$ versus stellar mass,  compared to observational data from DGS and KINGFISH. Simulated galaxies follow the observed trend with stellar mass in both cases.
We note that a few CLUES galaxies
 fall slightly below their observational counterparts at M$_{*}\approx$2-5$\times$10$^{8}$M$_\odot$, 
indicating that they host fewer  young stars. 
This may be due to environmental effects.

\begin{figure}
     \resizebox{\hsize}{!}{\includegraphics{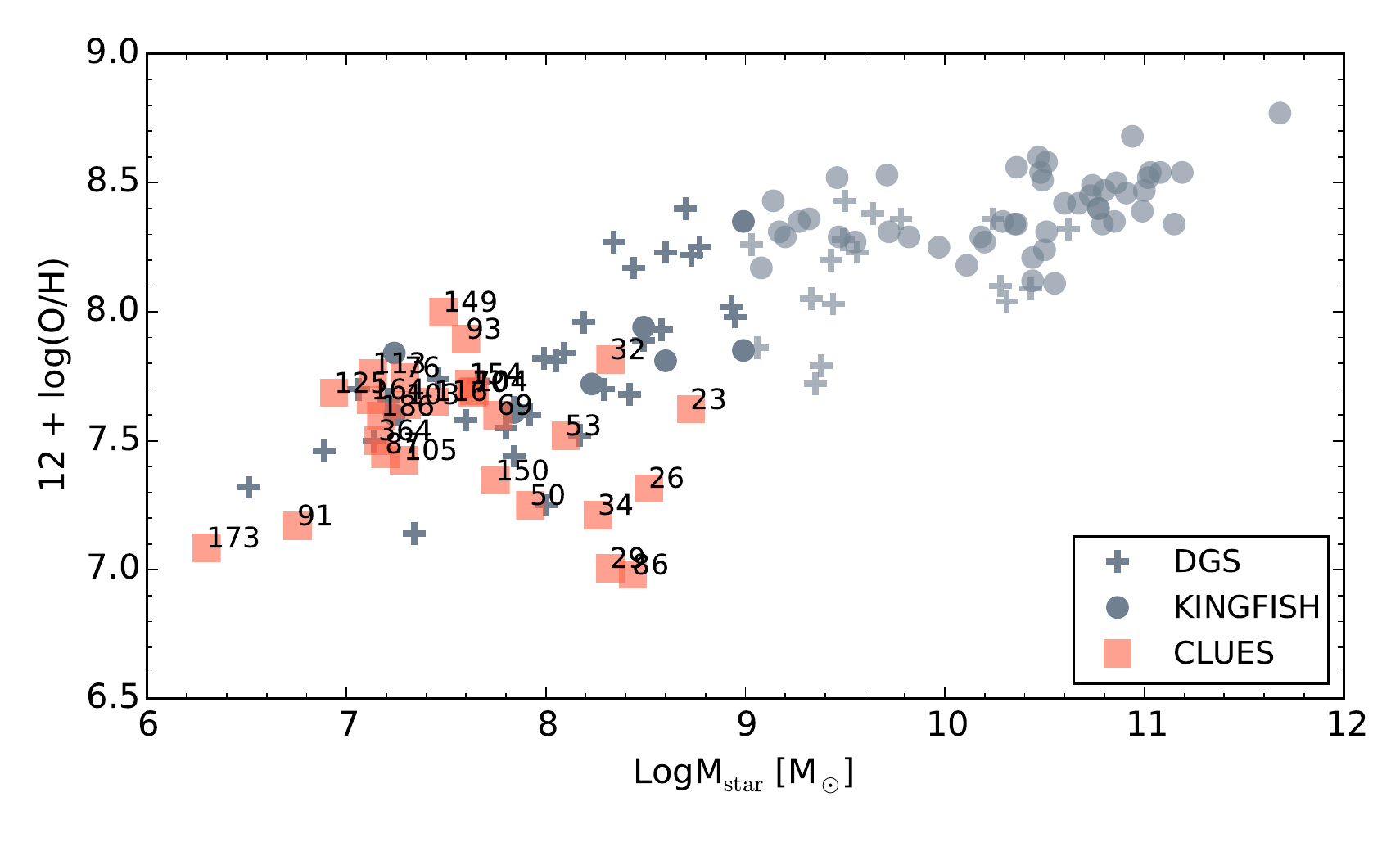}}
   \caption{Stellar mass compared to gas metallicity in terms of the oxygen abundance. The CLUES sample of star-forming dwarf galaxies lies in the low-metallicity region overlapping DGS data.}
     \label{Figmet}%
    \end{figure}

\begin{figure}
     \resizebox{\hsize}{!}{\includegraphics{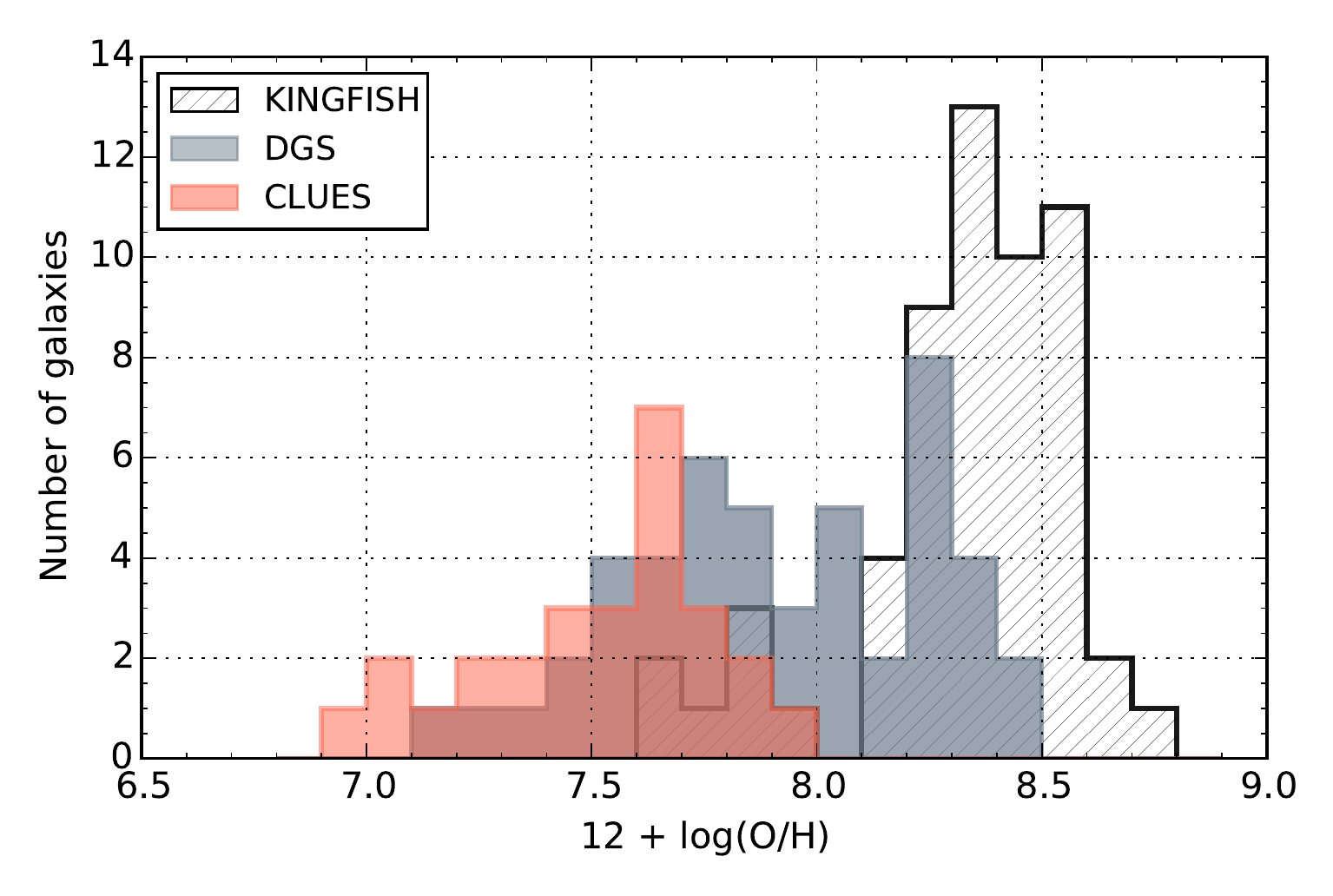}}  
   \caption{Metallicity distribution of the CLUES sample of star-forming dwarf galaxies, compared to that of the DGS and KINGFISH surveys.}
     \label{Figmethist}%
    \end{figure}


\subsection{HI and H$_2$ gas contents} \label{HIH2mass}

To ignore the spurious effects of hot ionized (HII) gas, we only provide GRASIL-3D with gas particles with temperatures lower than 2$\times$10$^4$K. 
For the simulated dwarfs that contain hot gas (50\% of the star-forming sample),  this temperature  cut  cleanly separates the cold and hot gas phases in a temperature histogram.
Therefore we can safely assert that
 M$_{\rm cold gas}$=M$_{\rm HI}$+M$_{\rm H2}$+M$_{\rm He}$+M$_{\rm Z}$, where M$_{\rm He}$ and M$_{\rm Z}$ are the mass fractions in helium and in metals, respectively. 

Concerning atomic and molecular gas fractions, 
as explained previously, in GRASIL-3D the fraction of gas  in the form of molecular clouds $f_{mc}$ is derived under the assumption that MCs are defined by a density threshold, $\rho_{mc,thres}$, and by a probability distribution function (PDF) for the gas distribution at subresolution scales with dispersion $\sigma$.
 When the MC mass is known,
  HI and H$_2$ masses are derived by  correcting for the fraction of mass in helium  and
  ignoring the fraction of mass in metals.

We show in Fig. \ref{FigHcontent} the HI and H$_2$ gas contents of 
the simulated galaxies
 compared to observational data inside the dwarf stellar mass range.
The upper panel proves that the fraction of HI gas mass per total stellar mass in simulated galaxies agrees with observational data. 
Regarding the fraction of molecular gas H$_2$, the middle and lower panels show observational data assuming a metallicity-dependent and a  constant-Galactic CO-to-H$_2$ conversion factor, respectively.
The gas density threshold of reference used in this work ($\rho_{mc,thres}$=3.3$\times$10$^{9}$ M$_\odot$ kpc$^{-3}$) provides an acceptable match to the observational M$_{\rm H2}$ data resulting from assuming a  metallicity-dependent conversion factor (middle panel); however, a value one order of magnitude higher  is required to fit observed data that assume a constant conversion factor (lower panel).
This translates into a decrease of M$_{\rm H2}$ in CLUES galaxies of an average factor of $\approx 4$, as can be observed by comparing the middle and lower panels of Fig. \ref{FigHcontent}.

In the following sections, the infrared luminosity results are shown for these two different gas density threshold values. 
As we show, qualitatively the final conclusions of this study are unchanged 
when using either of these two $\rho_{mc,thres}$ parametrizations.
We therefore do not attempt to
complicate the picture by
considering different methods for estimating the molecular gas fraction or taking  physical processes into account that may affect the formation of H$_2$ on dust grains (such as interstellar radiation fields or dust density).

\subsection{Gas metallicity}

 Among other metals, GASOLINE  traces the evolution of oxygen in gas and stellar particles, which is produced as a yield of SNII and SNIa explosions (see Sect. \ref{gasoline}).
 The 12+log(O/H) metallicity for each object was computed by averaging the oxygen mass fraction from cold gas particles inside 2$\times$R$_{\rm half}$, where R$_{\rm half}$ is the median radial distance of gas particles from the center of the galaxy. 

We show the mass-metallicity relation of CLUES galaxies
in Fig. \ref{Figmet}. In general, they follow the observational relation, although a few of them with LogM$_{*} \sim$8.4 have lower metallicities than the data.
Furthermore, Fig. \ref{Figmethist} displays the metallicity distribution.
CLUES dwarfs are certainly low-metallicity galaxies: their 12+log(O/H) values range between 7.0 and 8.0 dex, 
with a low-metallicity tail showing a very similar number distribution to that of DGS.

{
\subsection{Dust-to-gas mass ratio}\label{dgratio}
Figure \ref{DtoG_metmstar} presents the total (i.e., in a given galaxy) D/G ratio versus 12+log(O/H) gas metallicity colored by stellar mass. 
CLUES galaxies  at very low metallicities show dispersion in their dust masses, as is observed. 
We note, however, that all observed galaxies with metallicities below 12+log(O/H)$\lesssim 7.7$ (and many other at higher metallicities) have both uncertain
i)  H$_2$ masses  \citep[the absence of data has been replaced by an M$_{\rm H2}$ correction][]{RR14}
and 
ii) dust masses, \citep[due to non-detections of the galaxy at wavelengths larger than 160 $\mu$m][]{RR15}.
More observational work is needed to surpass these uncertainties.
We recall that we have assumed a broken power-law dependence of D/G on Z metallicity (see Eq. \ref{dustfrac}) to calculate the CLUES dwarf dust masses.
The dust masses and the consequences on the final SEDs that result by assuming a linear dependence of D/G on Z instead are shown in Sect. \ref{sec:varDG}. 
\\ 


\noindent
Based on this analysis, we can conclude that the properties of the sample of CLUES simulated star-forming dwarfs  are  consistent with observations, and we proceed to the SED  analysis. 

   \begin{figure}
     \resizebox{\hsize}{!}{\includegraphics{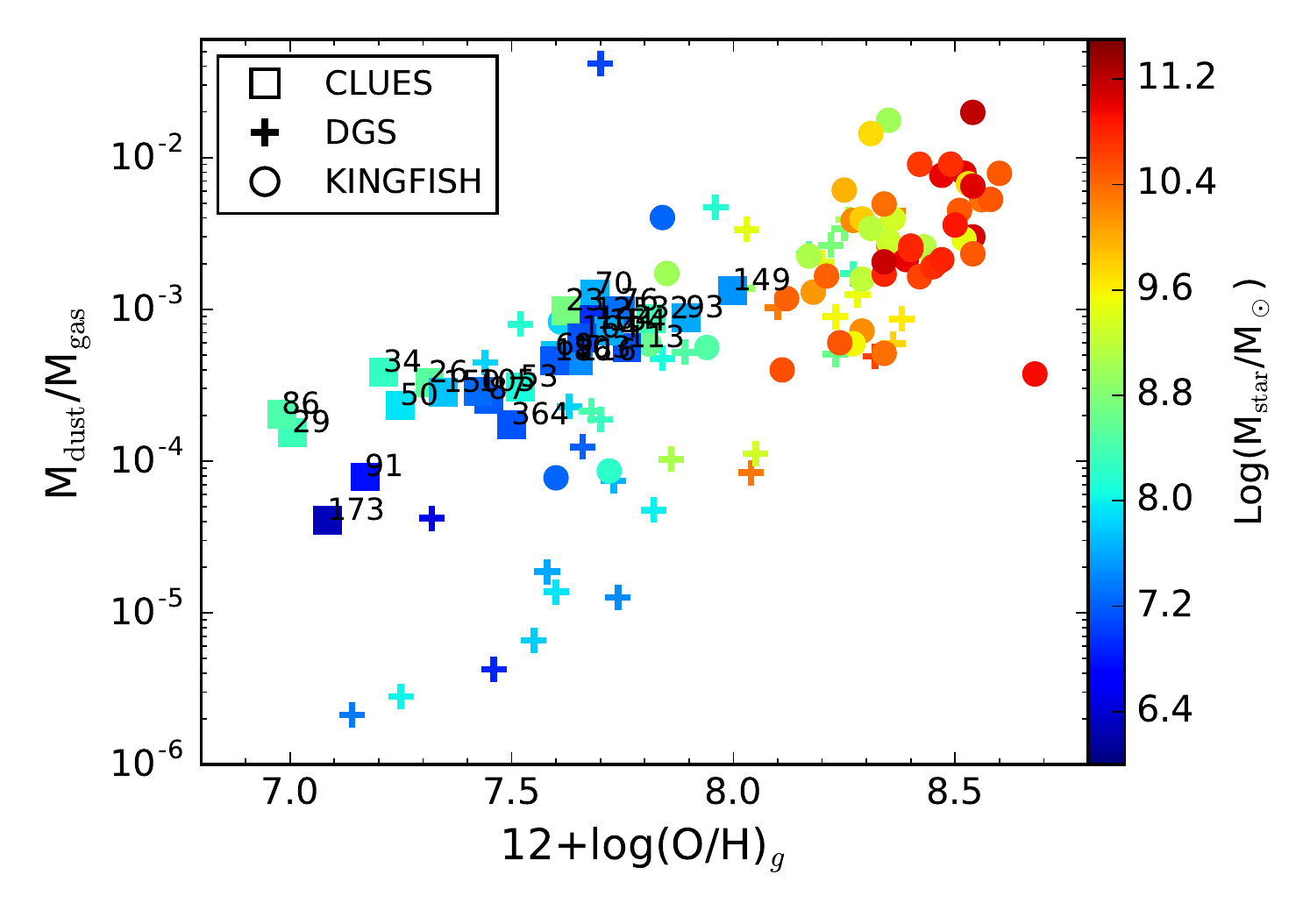}}
   \caption{
  Dust-to-gas mass ratio versus gas metallicity in terms of oxygen abundances. Data points are colored by stellar mass. 
   We assume in this study a broken power-law dependency between the D/G ratio and Z metallicity \citep{RR14} in each grid cell (Eq. \ref{dustfrac}).
Taking the uncertainty in the determination of both M$_{\rm H2}$ and M$_{\rm dust}$ for all observed galaxies with 12+log(O/H)$ \lesssim 7.7$  into account,
 we consider that this approach provides satisfactory estimates of the total dust masses of the CLUES dwarf sample as compared to DGS and KINGFISH.
   }
     \label{DtoG_metmstar}%
    \end{figure}

\section{Results: The diversity of emission in the IR-submm range} \label{sec:emi}

\subsection{SEDs}

To explore how CLUES star-forming dwarfs behave 
regarding their IR-submm emission, we show
in Fig. \ref{FigSEDs}  their SEDs as calculated with GRASIL-3D.
The intrinsic stellar emission is shown in red, the MC emission in blue, the emission from cirrus in green, and the black curve gives the angle-averaged total emission. 
The SEDs show a surprising variety, mimicking the different characteristics shown in DGS and HAPLESS SEDs as described by \citet{RR15} and \citet{Clark:2015}, respectively.

Specific examples  of SEDs with clearly defined patterns are 
i) IR peak broadening in 
numbers 34, 104, 149, and 164;
ii) submm excess and slope flattening in numbers 26, 32, 50, and 86;
and iii) low PAH emission in  
numbers 26, 32, 53, 116, and 125.
Other galaxies show less clear patterns and/or mixed properties, for example, 
numbers 23, 34, 154, and 93.
Finally, in three cases none of these features are apparent: numbers 103, 87, and 186.  

Contrary to observational data, where no SED decomposition into different dust components is directly available,
GRASIL-3D  provides  the MC and cirrus emissions. 
Fig. \ref{FigSEDs} shows that 
  two distinct dust components can be distinguished in the simulated galaxies: a colder (cirrus, green) and 
a warmer component (MCs, blue).\footnote{
We recall that the terms `cold' and `warm' refer to the location of the respective emission maxima
(sum of the emissions of grains of different types, sizes and temperatures),
 therefore they refer to a temperature that is only effective (see Sect. \ref{dustmodel}). }
}
In this way, a first answer to the question of the double dust component SED in dwarf galaxies
suggested in numerous studies
\citep[e.g.,][]{Bendo2010, Bendo2012, Bendo2012b, Boquien2011, RR13}
is naturally given by our modeling: the two components are the emission from 
PAH-depleted
dust in MCs that is heated by young stars, and from dust in the cirrus that is heated by more evolved stars.

Each of these two SED components is 
 shaped by their respective
  dust grain 
 total effective 
 temperatures
(determined by the location of  the emission maxima) and the flux intensities at their
respective maxima. 
The final IR-submm SED is the addition of these two (MC and cirrus) contributions.
According to the relative positions of the two maxima and the respective flux values there,
 a continuum of possibilities emerges from this combination, 
with extremal behaviors occuring when one component dominates over the other.
In the next subsections the implications 
of these edge situations and those of the intermediate possibilities
are described and discussed in detail.  
To better quantify the comparisons with observations,  this is done through color-color analyses.

\subsection{IR peak broadening} \label{sect:broad}
While the SEDs of larger and metal-richer galaxies peak at around 100-160 $\mu$m 
\citep[as most KINGFISH SEDs show, see Fig. 13,][]{RR15}, dwarf galaxies  in general present a broad IR SED with  a wide maximum. 
This behavior is reproduced in the GRASIL-3D SEDs of the CLUES dwarf galaxies. In particular, 
two different broadening features are visible in Fig. \ref{FigSEDs}, having an equivalent in the
SEDs of DGS galaxies
\citep[see Fig. 12,][]{RR15}.
\begin{enumerate}
\item
In some cases, the MC emission is much stronger (and effectively hotter) than that of the cirrus
in such a way that it hides PAH emission and produces a broad maximum that typically peaks in the  20 - 40 $\mu$m range. 
This is the case for  numbers 26, 32, 50, 116, and 164,
 and
to a lesser extent for  
numbers 86, 150, and 364.
A typical observational  counterpart is DGS galaxy HS0017+1055. 
\item
A different characteristic situation arises when both
the MC and cirrus emissions have similar intensities at their
(separated) maxima, typically at $\lambda \sim 40$ and $\sim 160\,\mu$m, respectively.
 In this case, the combined emission  is rather flat between the two maxima,
or with a slight slope if the maxima heights are somewhat dissimilar. 
Typical examples of this situation are  numbers 29, 53, 69, 104, 154, 93, 105, 113, and 125.
On the observational side,  DGS galaxy Pox186 is an example, and most SEDs of the HAPLESS galaxies 
\citep[Fig. A2 of][]{Clark:2015} could reflect this situation. 
\end{enumerate}

To quantify these behaviors, we show in 
Fig. \ref{FigBroad} the PACS/PACS  
  $\nu$L$_{70}$/$\nu$L$_{100}$ vs $\nu$L$_{100}$/$\nu$L$_{160}$ color-color diagram, 
  which traces the peak of the SED \citep{RR13}.
CLUES star-forming dwarfs (filled squares) are colored according to the amount of energy absorbed by their MCs per unit dust mass in MCs,
(E$_{\rm abs}$/M$_{\rm dust}$)$_{\rm MC}$
 (see colorbar).
We assume that the energy absorbed by MCs is equal to their infrared luminosity (i.e., energy emitted), which we compute by integrating the MC curve between 4 and 2000 $\mu$m.
In the upper and lower panels we compare 
CLUES dwarf galaxies
to the DGS (crosses) and KINGFISH (circles) samples, 
and in the middle panel 
we compare
to HAPLESS galaxies (triangles).
We show the HAPLESS
sample separately from DGS and KINGFISH because as explained before, these galaxies were selected for their
 high dust content and  are therefore a biased sample. 
The upper and lower panels differ in the $t_0$ value used.
Contrasting symbols represent galaxies
 within the $\rm 10^6<M_*(M_{\odot})<10^9$  stellar mass range, while faint symbols stand for
more massive galaxies. 
The two columns in the figure show results using different values for the gas density threshold that determines the molecular gas mass $\rho_{mc,thres}$, 3.3$\times$10$^9$ or 3.3$\times$10$^{10}$ M$_\odot$ kpc$^{-3}$. As explained in Sect. \ref{HIH2mass}, the first matches observational data assuming a metallicity-dependent CO-to-H$_2$ conversion factor, while the higher parameter is required for fitting H$_2$ masses calculated assuming the constant Galactic value.

The upper panels show
 most  KINGFISH high-mass galaxies gathered  in the lower left corner of the figure where $\nu$L$_{70}$/$\nu$L$_{100}<$1 and   $\nu$L$_{100}$/$\nu$L$_{160}\sim$1, because they peak at high 
wavelengths.
A few CLUES dwarfs whose MCs are poorly heated by young stars
 (blue squares,
numbers 70, 76, 87, 103, and 186) have SEDs that also peak at  around 100-160 $\mu$m and thus appear in the same location.  

However, 
most of the simulated, DGS and KINGFISH galaxies 
inside the "dwarf" stellar mass range, show a different behavior.
On one hand,
those that show a maximum emission in the 20 - 40 $\mu$m range 
(simulated dwarfs with dominant MC emission) 
consequently have $\nu$L$_{70}$/$\nu$L$_{100}>$1 and $\nu$L$_{100}$/$\nu$L$_{160}>$1, which places them in the upper right part of the diagram.
We note that DGS galaxies whose SED data have been fit in \citet{RR15} with an additional emission component as a MIR modified black body (marked with open circles) appear in this same area. 
On the other hand,
galaxies
showing an almost flat behavior in the $\sim 40-160 \,\mu$m range of their SEDs
are clustered closer to the $\nu$L$_{70}$/$\nu$L$_{100}= \nu$L$_{100}$/$\nu$L$_{160} =1$ position of the diagram.

To reinforce our interpretation that the warmer effective temperatures of MC dust grains in CLUES star-forming dwarfs 
are due to the higher energy absorption of
this dust (per unit mass), we decreased the energy input from young stars to  $t_0 = 5$ Myr, and 
 repeated
the SED calculation.
The results are shown in the lower panels of  Fig. \ref{FigBroad}.
As expected, energy absorption by MCs  decreased for most simulated galaxies, in some cases to the extent that 
their MC emission faded away 
to reveal the cirrus component
(dark blue squares). 
As a result,  they gather where the massive KINGFISH galaxies lie. 

We therefore postulate that IR peak broadening in DGS galaxies, regardless of the peak position, 
 has the same origin: the multi-temperature warm dust emission in MCs, heated by young stars. 
The two broadening types come from the 
relative intensity of the MCs and the cirrus maxima. The stronger the MC emission
 (we recall that the maxima appears in the 20 - 40 $\mu$m range), the more displaced
toward the right upper corner  in the
$\nu$L$_{70}$/$\nu$L$_{100}$ vs $\nu$L$_{100}$/$\nu$L$_{160}$ color-color diagram, tracing the peak of the SED.

In the middle panels of Fig.  \ref{FigBroad} we compare our results to 
HAPLESS\footnote{Note that here 
the available data are L$_{60}$ instead of L$_{70}$.} dwarf galaxies, 
which are biased against low dust mass content.
These galaxies therefore tend to have low values of the energy absorbed by MCs
per unit dust mass, $(E_{\rm abs}/M_{\rm dust})_{\rm MC}$.
Their SEDs are not expected to show MC emission
dominance, but rather broad maxima typically in the range between $\lambda \sim 40$ and $\sim 160\,\mu$m, as is the case.

Finally, we compare the CLUES results shown in the left and right columns of Fig. \ref{FigBroad} for different values of $\rho_{mc,thres}$.
 The only difference is observed in the color of the representative points, 
 which take higher values of (E$_{\rm abs}$/M$_{\rm dust}$)$_{\rm MC}$ 
 in the case of an approximately
 four times 
 lower M$_{\rm H2}$
($\rho_{mc,thres}$=3.3$\times$10$^{10}$, right column),
 as expected: a lower molecular gas mass (which implies a lower amount of dust) absorbs the same energy coming from young stars.
Nonetheless, the general arrangement of the points across the diagram does not vary,
and therefore
our conclusions remain valid.


\subsection{Submillimeter excess and slope flattening}
\label{Submm-excess}

Figure \ref{FigExcess} shows the PACS/SPIRE 
$\nu$L$_{100}$/$\nu$L$_{250}$ vs $\nu$L$_{250}$/$\nu$L$_{500}$ color-color diagram.
The symbol and color-coding are the same as in Fig. \ref{FigBroad}.

This diagram reflects the variation in emissivity index $\beta$, which is
understood as a measure of the final FIR/submm slope of the SED (see Eq. \ref{ModBB}). 
Physically, it is an intrinsic optical property of grains: a typical value of $\beta$=2.0 is found for Galactic grains and is commonly used to model the SEDs of high-mass galaxies, showing a good agreement with data.
In contrast, dwarf galaxies have been reported to present flatter slopes, reflecting an excess of emission at submm wavelengths.

In this study, both the MC and cirrus components were modeled assuming $\beta$=2.0, 
but their combination yields a final effective slope --or $\beta_{\rm eff}$-- that in general departs from this value.
Figure  \ref{FigExcess} shows three curves depicting the theoretical \textit{Herschel} luminosity ratios obtained assuming modified black bodies of different fixed emissivity indices, which we accordingly also call $\beta_{\rm eff}$, drawn from Fig. 10 in \citet{RR13}.

While the bulk of
 high-mass galaxies
follow the higher $\beta_{\rm eff}$
(i.e., with no submm excess found),
 the observed dwarf galaxies present lower $\nu$L$_{250}$/$\nu$L$_{500}$ ratios, occupying  
the $\beta_{\rm eff}<1.5$ region.
We note in particular the presence of the DGS galaxies marked in \citet{RR13} as having  an excess of emission at 500 $\mu$m 
with respect to modified black body $\beta=2$ fits to the data
(open squares).
This indicates a general flattening of the submm SED slopes of these galaxies as compared to the SEDs of more massive and metal-rich galaxies. 
This same behavior is also observed in the SEDs of the CLUES star-forming dwarf galaxies.

When we analyze the figure in more detail,
we see that CLUES dwarfs with an IR slope flattening caused by a dominant MC emission 
(i.e., numbers 26, 32, 50, 116, 164,  86, 150, and 364)
appear between the $\beta_{\rm eff}=1.0$ and $1.5$ lines and toward the upper right corner,
where the inequalities 
 $\nu$L$_{100} \gg \nu$L$_{250}$
 and $\nu$L$_{250} \gg \nu$L$_{500}$ clearly hold.
On the other hand, when IR peak broadening is instead caused by a similar MC and cirrus peak intensity, 
the $\nu$L$_{100}$ luminosity is similar to, and in most cases only slightly higher than, $\nu$L$_{250}$,
 placing these cases close to the dotted line that represents the
  $\nu$L$_{100}$= $\nu$L$_{250}$
  equality. 
Finally, CLUES galaxies with (almost) bare cirrus emission (dark blue squares, numbers 103, 76, 87, and 186) 
are the closest to the dotted line and also have the lowest $\nu$L$_{250}$/$\nu$L$_{500}$ ratios.

As in Fig. \ref{FigBroad},
the upper and lower panels of Fig. \ref{FigExcess} differ in the MC escape timescale used in the GRASIL-3D run:  $t_0=40$ Myr in the upper panels; $t_0=5$ Myr in the lower panels.
When we use this lower value, far fewer CLUES galaxies have stars young enough  to live in MCs, and thus 
 the energy absorbed is very low (dark blue, see color bar). 
 Their SEDs 
 change to roughly present only the cirrus dust component, steepening
 their submm slopes and bringing  
  their representative points
between the lines corresponding to the  $\beta_{\rm eff}=$1.5 and 2.0 fits, where massive
galaxies lie.
We conclude that the heating of MCs by young stars  drives 
the SED flattening around the 500 $\mu$m SPIRE band.

HAPLESS galaxies (middle panels of Fig. \ref{FigExcess}) show the same general behaviour as DGS or KINGFISH,
but we note that they tend to take lower  $\nu$L$_{100}$/$\nu$L$_{250}$ values than DGS and KINGFISH because they mostly present two  different  emission components \citep[see Fig. A2 of][]{Clark:2015}, and  it is rare for one of them to dominate over the other.

Comparing the two columns of Fig. \ref{FigExcess},
we again observe the subtle increase in
the $(E_{\rm abs}/M_{\rm dust})_{\rm MC}$  parameter
 (according to the color bar) when using  $\rho_{mc,thres}$=3.3$\times$10$^{10}$ M$_\odot$ kpc$^{-3}$, as a result of the lower MC dust mass.
Since a lower amount of MC dust grains absorbs the same energy coming from young stars, MCs acquire a higher global effective temperature. 
On the other hand, the cirrus phase has gained the difference in dust mass,
consequently decreasing its global temperature. These two temperature variations render a greater
 separation between both SED components. 
As a result, with a higher value of $\rho_{mc,thres}$, the emergence of an excess of submm emission thanks to the combined contribution of the emission of the two dust components is more 
likely.
This accounts for the slight displacement of points toward higher  $\nu$L$_{250}$/$\nu$L$_{500}$ values, which 
 is  noticeable (barely) only in the case of simulated dwarf galaxies with
similar MC and cirrus emission intensities. 
The global  differences in the PACS/SPIRE diagram are nonetheless insignificant, therefore we can conclude that the interpretation of our results is independent of the gas density threshold value used,
which translates in that it is independent of an 
 average factor of 4 higher or lower M$_{\rm H2}$ in the galaxy.
The effects of varying $\rho_{mc,thres}$ as well as the PDF dispersion $\sigma$ are  explored in detail in Sect. \ref{ParVar}.

\subsection{PAH emission}

\begin{figure}
    \resizebox{\hsize}{!}{\includegraphics{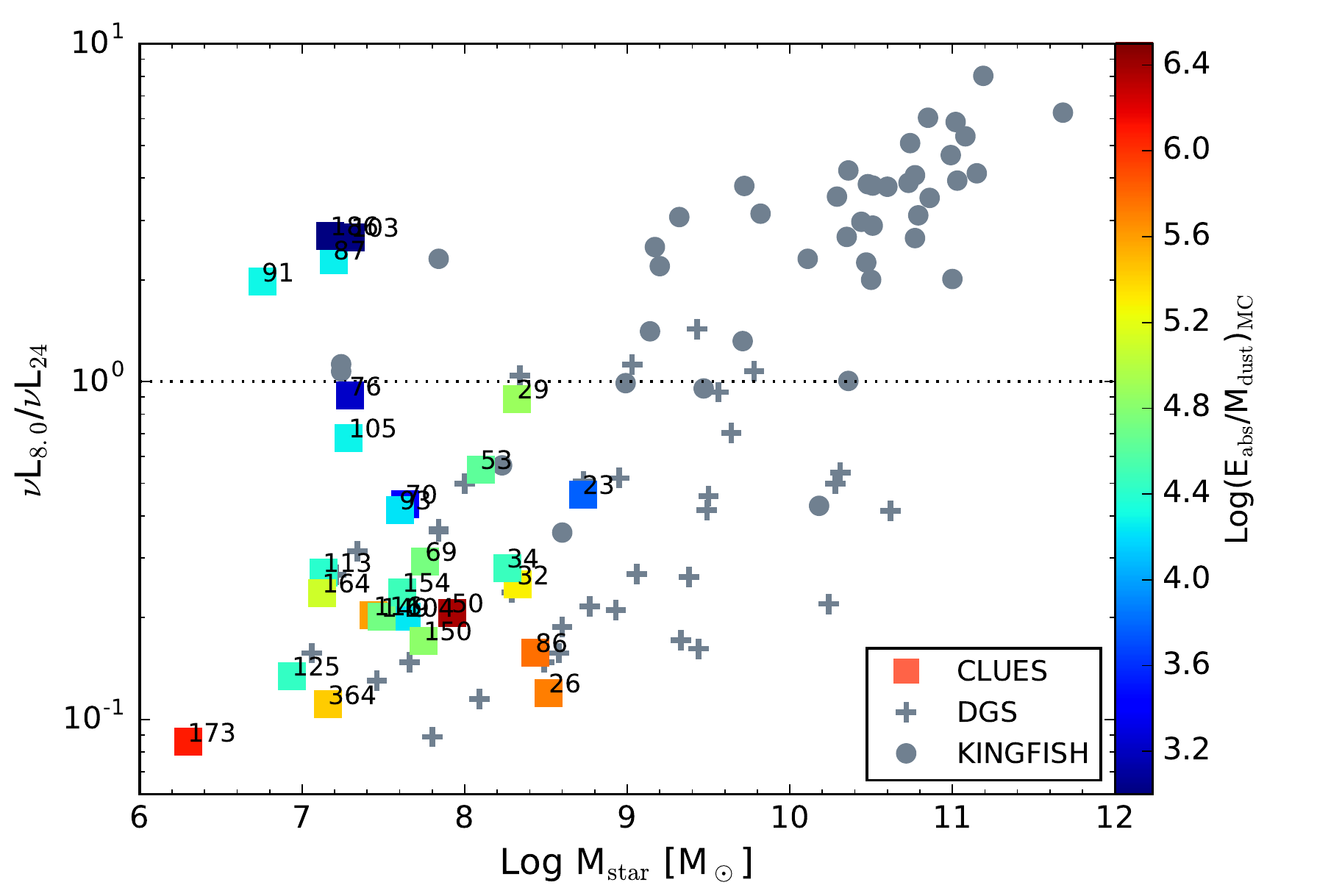}} 
   \caption{Stellar mass versus the $\nu$L$_{8.0}$/$\nu$L$_{24}$ ratio, which reflects the strengths of PAH emissions. DGS galaxies show \textit{IRAS} 22 $\mu$m data instead of MIPS 24 $\mu$m.}
              \label{FigPAH}%
    \end{figure}

The PAH band emission in dwarf galaxy SEDs appears to be very weak,
which has suggested the use of dust models with a low PAH abundance in order to explain the data.
Here we use one such model, characterized by $q_{\rm PAH}=$1.12\%,
 a suitable value for low-metallicity galaxies according to \citet{Draineb:2007} (see Sect. \ref{dustmodel}).

The $\nu$L$_{8.0}$/$\nu$L$_{24}$ ratio is a good estimator of the PAH 
peak intensities because
 it compares the strength of one of the most prominent peaks of the aromatic features (at $\lambda\approx8.6 \,\mu$m) and the luminosity  
immediately outside the PAH emission domain.
Figure \ref{FigPAH} shows this IRAC 8$\mu$m/MIPS 24$\mu$m \textit{Spitzer} luminosity ratio compared to stellar mass.
The majority of the CLUES star-forming dwarfs show very low PAH emission, the same as has been reported for DGS  \citep{RR15}. 
As we noted in Sect. \ref{sect:broad},
MC emission can i) completely hide PAH bands when the maximum of the MC emission is higher than the maximum of the cirrus component, 
or ii) partially hide PAH bands when both MC and cirrus maxima tend to have comparable strengths.
Only dwarfs 
number 91, 87, 186, 103, and 76 in Fig. \ref{FigPAH}
show a mid-infrared emission
where
$\nu$L$_{8.0}>\nu$L$_{24}$.
These are precisely the cases in which MC emission is poorer 
as a result of a low energy absorption from young stars
(dark blue squares in the upper panels of Figs. \ref{FigBroad} and Fig. \ref{FigExcess}). This situation prevents the concealment of PAH cirrus emission.

We propose that the weakness or lack of PAH emission in DGS galaxies 
could be caused by the combined contribution of 
i) a low PAH abundance and ii) warm dust emission from PAH-devoid MCs that partially or completely hides the cirrus PAH features.    

We  note that DGS galaxies show \textit{IRAS} 22 $\mu$m data instead of MIPS 24 $\mu$m. CLUES luminosities at 22 $\mu$m are similar to those at 24 $\mu$m, only differing for the few galaxies  mentioned
above,
  where L$_{22}>$L$_{24}$. 
We therefore expect the few DGS galaxies with $\nu$L$_{8.0}$/$\nu$L$_{22}>1$
to slightly shift upward if the L$_{24}$ data were available. This does not affect our results since these points
would still be consistent with the rest of data inside the same stellar mass range.
The rest of DGS galaxies would remain unaltered or even shift downward.


\section{Discussion} \label{sec:disc}

\begin{figure}
     \resizebox{\hsize}{!}{\includegraphics{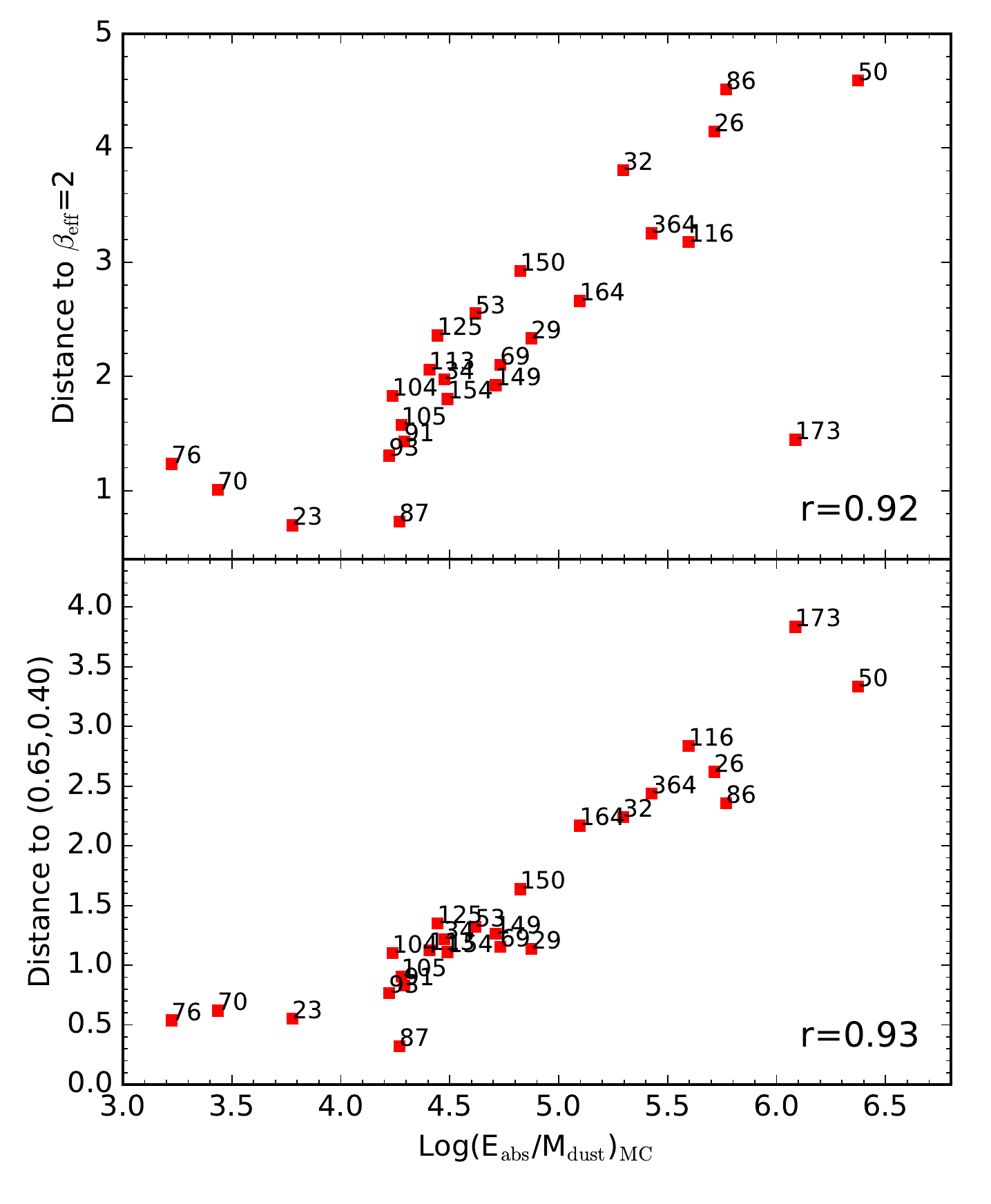}}   
   \caption{Correlations of the $(E_{\rm abs}/M_{\rm dust})_{\rm MC}$ ratio with distance to diagram positions representative of zero injection of energy into molecular clouds: the $\beta$=2 line in Fig. \ref{FigExcess} and the lower left corner of Fig. \ref{FigBroad}. Pearson correlation coefficients are given in each case.   
   }
              \label{Broad-Corr}%
    \end{figure}

\subsection{Nature of the heating engine } 

\label{Nature-cold-warm}

According to Fig. \ref{FigSEDs}, the cirrus components of the 
simulated dwarf galaxies  peak between $100 - 200 \,\mu$m,
wavelengths at which usually the total SEDs of more massive and metal-rich galaxies peak.
 Figure \ref{FigSEDs} also shows that the emission from MCs is effectively hotter than usual, showing in addition a wider diversity of shapes among the different galaxies.
In GRASIL-3D modeling, radiation 
 from evolved
stars heats the diffuse cirrus component where the bulk of dust resides.
In contrast,  MCs are heated by low $\lambda$ emissions from young stars
embedded in MCs.
Therefore the amount of energy young stars inject into MCs, $E_{\rm abs,MC}$, is expected to be a key parameter to understand 
the particularities in the IR-submm region of dwarf galaxy SEDs.
However, the MC grain temperature is most likely correlated to $E_{\rm abs,MC}$ per unit dust mass,
$(E_{\rm abs}/M_{\rm dust})_{\rm MC}$,
as we show below.
For this reason, simulated results are color-coded by this quantity in Figs. \ref{FigBroad}, \ref{FigExcess}, and \ref{FigPAH}.

In Fig. \ref{Broad-Corr} we show a measure of 
the 
energy injection effects
by representing the correlation between the $(E_{\rm abs}/M_{\rm dust})_{\rm MC}$ parameter and the 
distance to points  of zero injection in Figs. \ref{FigBroad} and \ref{FigExcess}. 
These points are the particular locations in each color-color diagram where galaxies lie when they are scarcely heated by young stars (appearing in dark blue color): the lower left corner of Fig. \ref{FigBroad} and  the $\beta_{\rm eff}$=2 curve in Fig. \ref{FigExcess}.

The upper panel of Fig. \ref{Broad-Corr} shows the distance from each CLUES  dwarf 
in Fig. \ref{FigExcess} 
to the 
$\beta_{\rm eff}$=2 curve. 
Simulated dwarfs with the maximum 
$(E_{\rm abs}/M_{\rm dust})_{\rm MC}$ ratio are farther away from 
this line than those with lower ratio values, 
the correlation having a Pearson coefficient of
 r=0.92. 
When we only use $E_{\rm abs, MC}$,
that is, when we do not divide by the total MC dust mass,
the correlation decreases to r=0.83.
The lower panel of Fig. \ref{Broad-Corr}   shows the distance from the position of each simulated dwarf in Fig. \ref{FigBroad}  to the 
lower left corner of this diagram (approximately (0.65,0.40)),
and again a clear correlation is visible,
with 
r=0.93.
When only $E_{\rm abs,MC}$ is used, the scattering increases 
and yields r=0.79.
We 
finally note that
although not depicted, concerning PAH emission,  low $(E_{\rm abs}/M_{\rm dust})_{\rm MC}$ ratio values are  correlated to high PAH emission, as can be deduced from Fig. \ref{FigPAH}.
CLUES dwarf galaxies with low or very low PAH emissions correspond to
high $(E_{\rm abs}/M_{\rm dust})_{\rm MC}$ ratios.

Our findings are independent of the $\rho_{mc,thres}$ value:
with the high-density threshold value ($\rho_{mc,thres}=3.3 \times 10^{10}$ M$_\odot$kpc$^{-3}$) the same clear correlations appear, improving when $E_{\rm abs}$ is taken by unit dust mass.  

These results corroborate that the simulated galaxies' global MC effective temperature,
which
defines the maxima location and
 drives the variability in the final IR-submm SED shapes,
  is adequately captured by the $(E_{\rm abs}/M_{\rm dust})_{\rm MC}$ ratio.

\subsection{Dependence of the results on the dust-to-gas mass assignation versus Z}
\label{sec:varDG}
As described in Sect. \ref{DGratio},
for this particular study on dwarf galaxies we have assumed in GRASIL-3D a 
local
dependence of the dust-to-gas mass ratio on metallicity following the broken power-law proposed in Table 1 of
 \citet{RR14}.
This scaling
consists of two power-laws, the low-metallicity end 
having a steeper slope that provides
 a lower total dust mass to such environments, as 
suggested   observationally and by dust evolution models.
For metal-richer galactic regions, in particular with metallicities
12+log(O/H)$\gtrsim8.0$,
the  standard convention of scaling the D/G ratio linearly with metallicity \citep["reference scaling" in][]{RR14} has been adopted, which is  long known to satisfactorily fit data 
of normal metal-richer galaxies.

\begin{figure}
     \resizebox{\hsize}{!}{\includegraphics{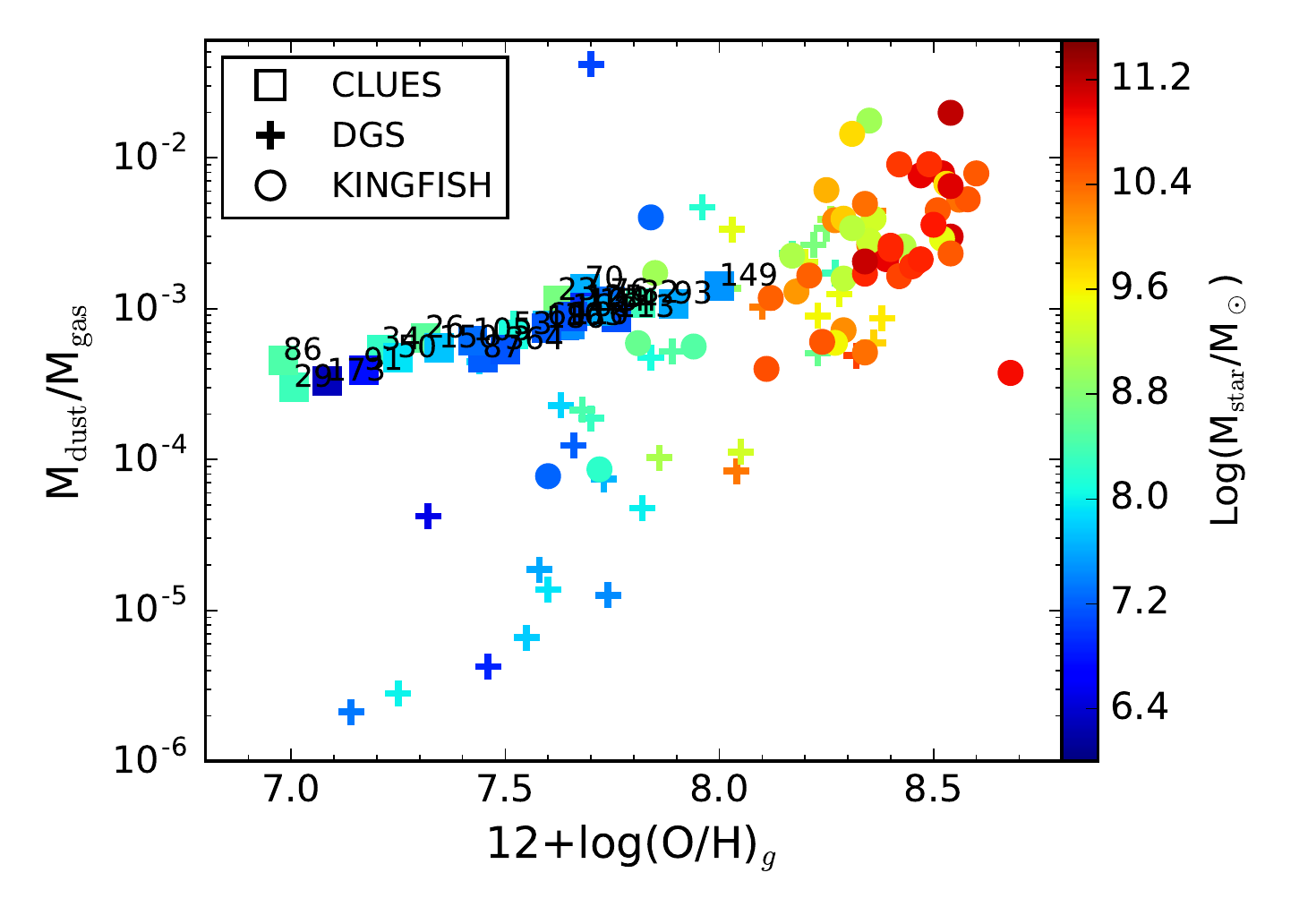}}
   \caption{
   Dust-to-gas mass ratio versus gas metallicity in terms of oxygen abundances. Data points are colored by stellar mass. A linear scaling is assumed between the D/G mass ratio and the metallicity inside each grid cell.
   }
              \label{Figdglinear}%
    \end{figure}

We  computed the dust masses, SEDs, and IR-submm luminosities that result from assuming
this reference scaling at all metallicities. 
Figure 
\ref{Figdglinear}
 shows the dust-to-gas mass ratios obtained in this case.
Galaxies   with 12+log(O/H)$\lesssim8.0$ show no dispersion and 
 end up with an excessive amount of dust as compared to data. 
However, the SEDs obtained with this dependence show that eventually this does not translate in a great difference in the final luminosities. 
In fact, only a few galaxies that have particularly very low global metallicities show perceptible changes in their SEDs (presenting a slightly stronger MC emission due to the now higher dust content), while the rest stay nearly the same.
The reason for this low variation of the final SED
is that
the grid cells that contribute the most to the emission 
of a given galaxy 
 are those
 with a higher dust content and thus a higher metallicity.

\subsection{Dependence of the results on the dust model for cirrus}
\begin{figure}
    \resizebox{\hsize}{!}{\includegraphics{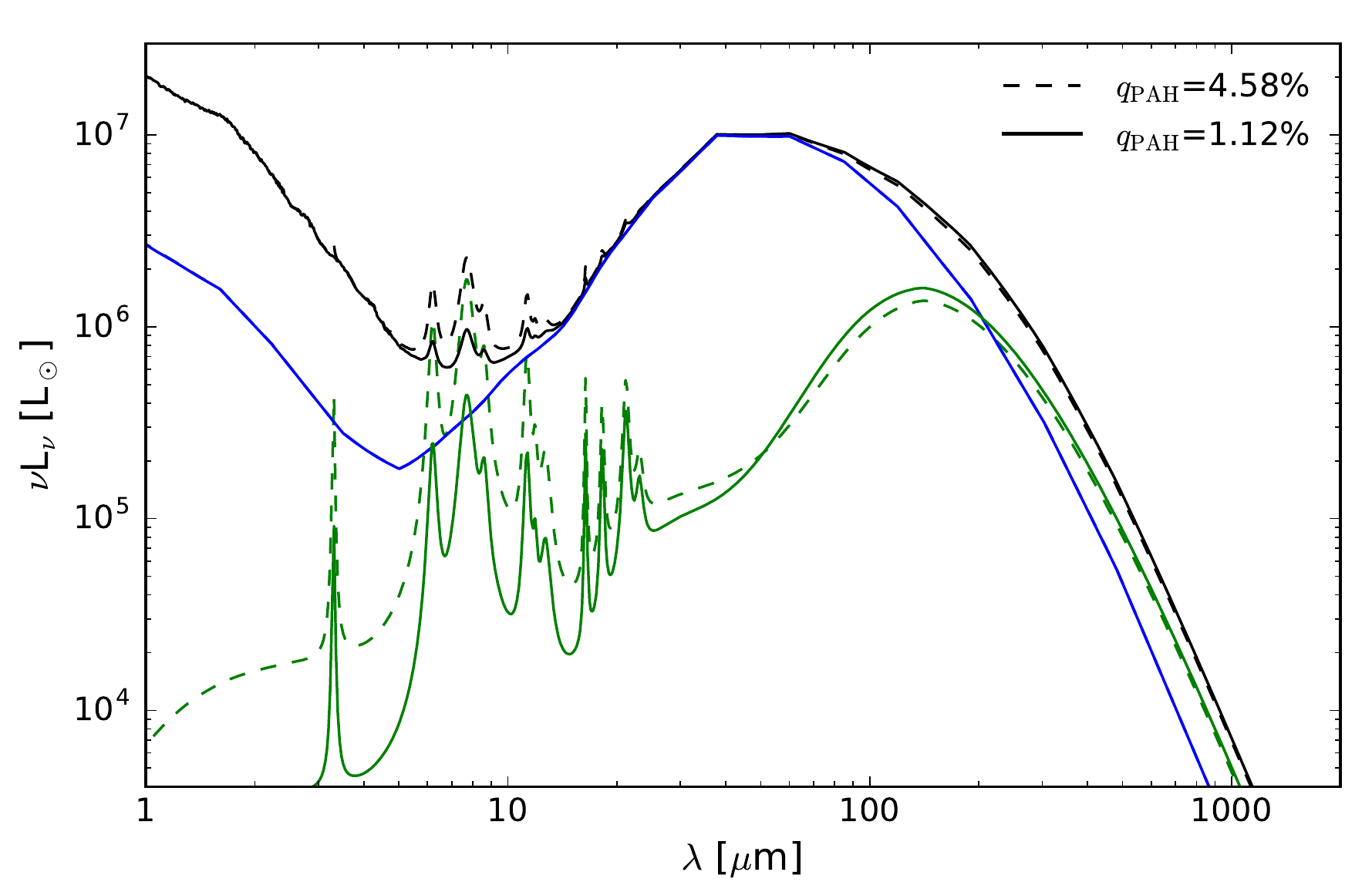}} 
   \caption{
   Comparison of SEDs of CLUES galaxy number 150 using different values  for the PAH abundance or PAH index $q_{\rm PAH}$ appropriate either for the Galaxy (4.58\%) or for low-metallicity galaxies (1.12\%).
   }
              \label{Figvarqpah}%
    \end{figure}

Spectra of low-metallicity galaxies show extremely low PAH band emission as compared to more metal-rich galaxies.
The fact that the most prominent peaks in the mid-infrared region are due to single-photon heating of the smallest PAH grains
 has led modelers to consider the use of different dust models containing fewer very small PAH grains in order to match the data. 
 In particular, \citet{Draineb:2007}
proposed seven different models that are  
 characterized by a different PAH index $q_{\rm PAH}$ each, ranging from 0.47\% to 4.58\%, where lower values are more appropriate for low-metallicity galaxies, and the highest value corresponds to a Galactic abundance.
In Fig. \ref{Figvarqpah} we show the differences in the SED of galaxy number 150 (and specifically in its cirrus emission) between adopting the standard Galactic value of 4.58\% and the value of reference used in this work, 1.12\%. 
PAH bands in the latter case are indeed mitigated, while with a high $q_{\rm PAH}$ they gain clear visibility.
We note that the effect would be much weaker in the case of dwarfs with dominating MC emission in the MIR region.

\subsection{Dependence of the results on parameter choices} 
\label{ParVar}

 Table 2 of DT14 shows the combinations of GRASIL-3D parameters used for their study of normal disk-like galaxies. All values proposed are within  observationally measured ranges. Here we  discuss the variation of
\begin{itemize}
\item  $t_0$, the escape timescale from MCs;
\item $\rho_{mc,thres}$ and $\sigma$, the gas density threshold and the parameter that governs the log-normal PDF function to calculate the gas fraction in MCs; and
\item $\tau_{mc}$, the optical depth of MCs, which is dependent on the values of the mass and radii of single molecular clouds ($m_{mc}$ and $r_{mc}$).
\end{itemize}

\subsubsection{Escape timescale from molecular clouds $t_0$}

\begin{figure}
     \resizebox{\hsize}{!}{\includegraphics{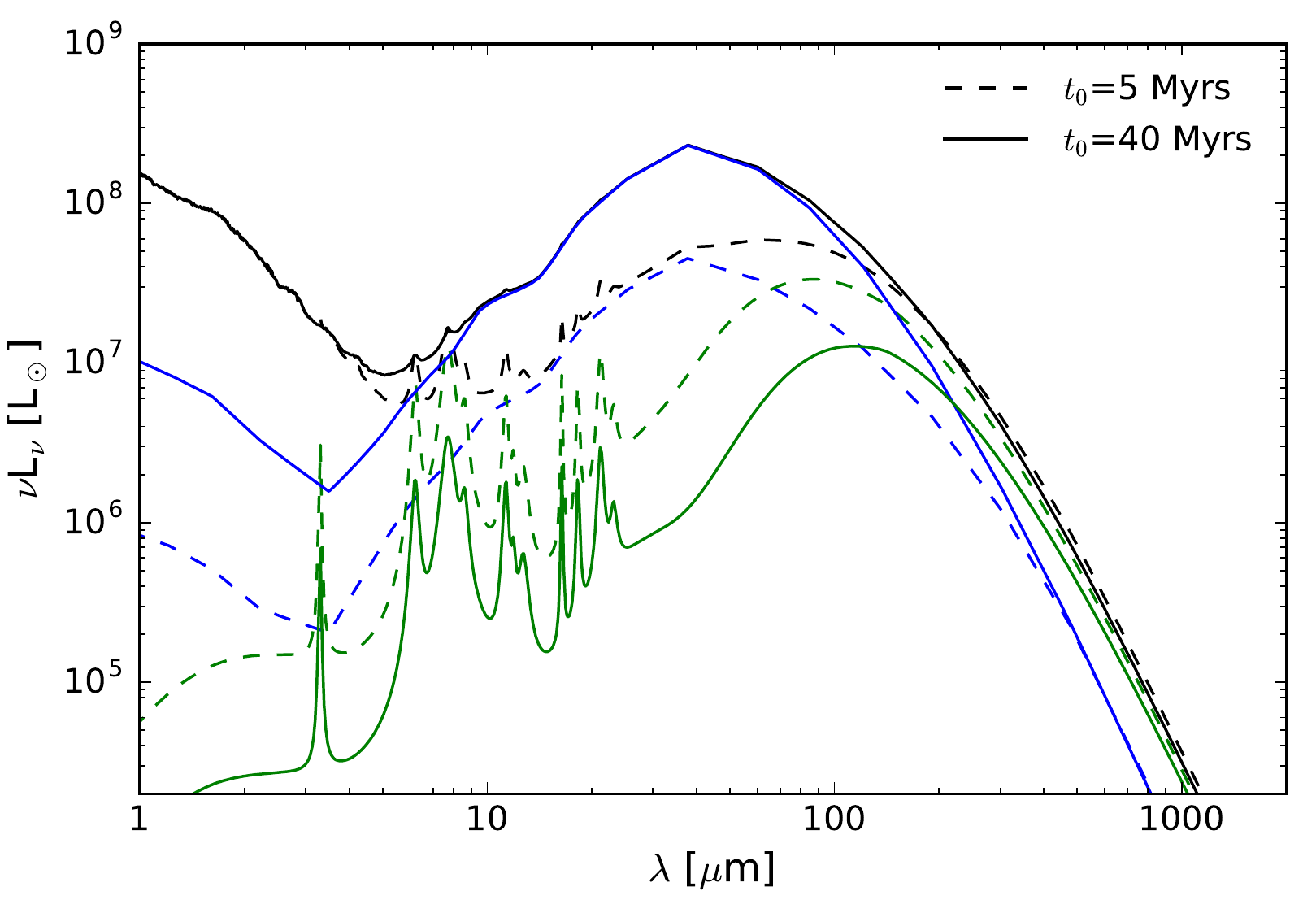}}
   \caption{Comparison of SEDs of CLUES galaxy number 26 using different values for the $t_0$ GRASIL-3D parameter.
   }
              \label{FigVart0}%
    \end{figure}

To vary $t_0$ means
to vary the amount and spectral distribution of the stellar energy that heats the MCs.
We show in Fig. \ref{FigVart0} the SEDs obtained for galaxy number 26 with $t_0$=5 Myr and $t_0$=40 Myr.
The higher the $t_0$ value, the higher the fraction of the total available stellar energy that heats the MCs.
Since the mass of MCs remains the same but now it absorbs more energy, the global effective MC grain temperature increases. As a result, the peak of the MC component broadens in the mid-infrared region, covering the possible PAH cirrus emission. 
With a low value of $t_0$, the percentage of galaxies in our sample with SEDs presenting the characteristic features observed in DGS, HAPLESS, and dwarf KINGFISH galaxies 
(i.e., IR peak broadening,  submm slope flattening, and low PAH emission)
decreases drastically, as has been shown before in the lower panels of Figs. \ref{FigBroad} and \ref{FigExcess}.
Within GRASIL-3D modeling, 
high $t_0$ values are therefore needed to account for dwarf SED properties. 
This is consistent with results of  S98 and S99, who showed that a high $t_0$ value is necessary to reproduce the SEDs of star-bursting galaxies.

\subsubsection{MC density threshold  $\rho_{mc,thres}$ and PDF dispersion $\sigma$}

This pair of parameters control the cirrus (HI) and MC (H$_2$) masses, regardless of the value of the MC escape timescale $t_0$.
We have shown the subtle consequences of increasing $\rho_{mc,thres}$ in Figs. \ref{FigBroad} and \ref{FigExcess}, which can be summarized in a slight additional warming of the MC component 
(per unit MC dust mass) as a result of a lower MC dust mass.
 In this section we show specific examples of the different SEDs obtained with different parametrizations.

\begin{figure*}
     \resizebox{\hsize}{!}{\includegraphics{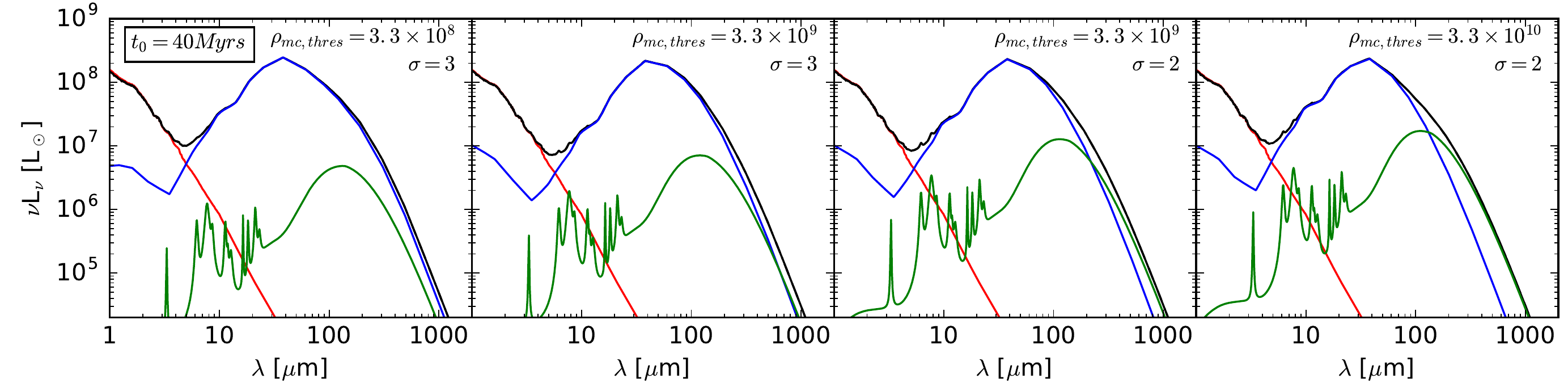}}
   \caption{Comparison of SEDs of galaxy number 26 obtained with different values of the density threshold $\rho_{mc,thres}$ and the PDF dispersion $\sigma$.}
              \label{FigVarPDF}%
    \end{figure*}

Maintaining a constant energy injection,  $t_0=40$ Myr,
we compare in Fig. \ref{FigVarPDF} four combinations of these parameters  ($\rho_{mc,thres}$ [M$_{\odot}$ kpc$^{-3}$], $\sigma$), in particular, 
the evolution in the SED of galaxy number 26 
when using the following pairs of values:
 (3.3$\times$10$^8$, 3), (3.3$\times$10$^9$, 3), (3.3$\times$10$^9$, 2), and (3.3$\times$10$^{10}$, 2). 
This evolution
from left to right
 corresponds to a  progressive MC (cirrus)  mass decrement (increment)
inducing a global increment (decrement) of MC (cirrus) dust temperature. 
Both effects produce  a gradual separation of the two dust components,
which gives rise to a broader total IR peak and to a flatter submm slope.

In particular, for
a fixed value of $\sigma$=3, the MC mass of galaxy number 26 is reduced by a factor of 1.4 from $\rho_{mc,thres}$=3.3$\times$10$^8$ M$_{\odot}$ kpc$^{-3}$ to 3.3$\times$10$^9$ M$_{\odot}$ kpc$^{-3}$, while with $\sigma$=2, the MC mass is reduced by a factor of 3.8 from $\rho_{mc,thres}$=3.3$\times$10$^9$ M$_{\odot}$ kpc$^{-3}$ to 3.3$\times$10$^{10}$ M$_{\odot}$ kpc$^{-3}$.
We found similar results  for all the simulated dwarfs in the sample.
Furthermore, the central panels of Fig. \ref{FigVarPDF} show that the effects of reducing the PDF dispersion $\sigma$  from 3 to 2 are qualitatively the same (and quantitavely  similar) to increasing the density threshold 
one order of magnitude: MC mass and emission decreases (by a factor of 2.8 for galaxy number 26) while the cirrus mass and emission increases. 
We note that although the SED of galaxy number 26 with $\rho_{mc,thres}$=3.3$\times$10$^8$ M$_{\odot}$ kpc$^{-3}$ seems to satisfy all observed dwarf SED characteristics, the eventual H$_2$ mass is at the limit of agreeing with observational data using a metallicity-dependent CO-to-H$_2$ conversion factor, and it does not agree with data using a constant factor (see the middle and lower panels of Fig. \ref{HIH2mass}). 
If the remaining galaxies in our sample change by a factor similar to galaxy number 26, 
not only would it lead to an excess of molecular gas, but to a great shortage of HI neutral gas. Therefore we do not consider this low $\rho_{mc,thres}$=3.3$\times$10$^8$ M$_{\odot}$ kpc$^{-3}$ value in our study.

\subsubsection{Masses and radii of molecular clouds}

The values of $m_{mc}$ and $r_{mc}$ are constrained observationally to range typically between 10$^5$-10$^6$ M$_\odot$ and 10-50 pc for normal spiral galaxies, as discussed in S98, S99, and DT14. 
 We have studied the impact on the SED of varying 
the mass of single MCs within its allowed range, while fixing the radius at 14 pc. This approximation is consistent since it its their combined response  ($m_{mc}/r_{mc}^2$) what matters, affecting the total emission through the optical depth of MCs $\tau_{mc}$.

\begin{figure*}
      \resizebox{\hsize}{!}{\includegraphics{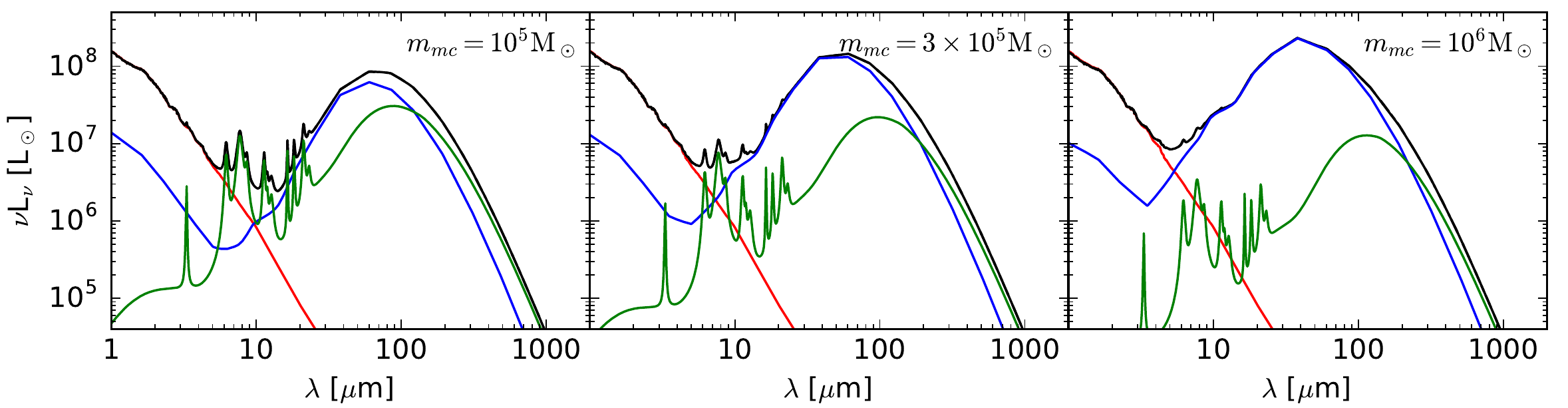}}  
   \caption{Comparison of SEDs of CLUES galaxy number  26 using different values of the mass of a single molecular cloud $m_{mc}$.}
              \label{FigVarMMC}%
    \end{figure*}

\begin{figure*}[h]
     \centering
   {\includegraphics[width=16 cm]{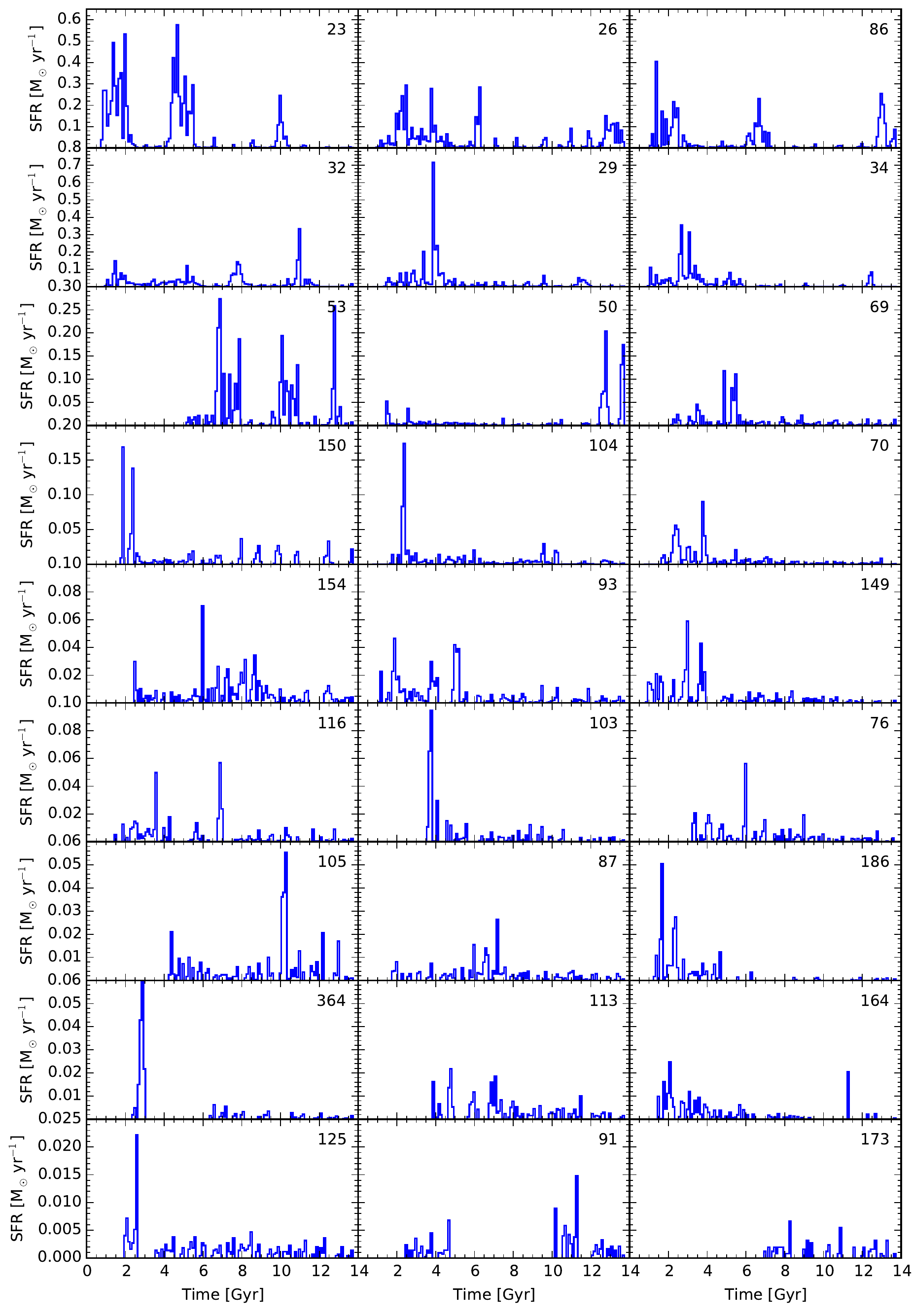}}
   \caption{Star formation histories of the CLUES sample of local star-forming dwarf galaxies ordered by decreasing stellar mass (from left to right and top to bottom).}
              \label{FigSFH}%
    \end{figure*}

We  calculated the SEDs of galaxy number 26 taking $m_{mc}$=10$^6$, 3$\times$10$^5$ and 10$^5$ M$_\odot$, and the rest of the parameters as usual.
Figure \ref{FigVarMMC} shows that 
as $m_{mc}$ decreases, $\tau_{mc}$ decreases (which means MCs absorb less energy per unit length).
With extremely low $m_{mc}$
a situation can be reached
 where UV photons escape from the MCs to the ISM, heating the cirrus.
This would occur at the expense of the MIR emission of MCs, which  gradually become unable to hide the
already low
 PAH
 emission from the cirrus. 
In addition, in this case, the
emission and temperature of the cirrus would increase, favoring PAH band detectability.
We therefore
 used   $m_{mc}$ $=$10$^6$ M$_\odot$ here 
to account for dwarfs with a very broad IR peak in the 20 - 40 $\mu$m range
and no detected PAH emission features.

\section{Summary and conclusions} \label{conclu}

The IR-submm emission of low-metallicity  dwarf galaxies considerably differs from that of more metal-rich dwarf galaxies.
Signatures of warmer dust in dwarf galaxies were first found with \textit{IRAS} and were later on confirmed with \textit{Spitzer}. 
More recently, an important step forward in the analysis of their IR-submm emission was made thanks to
the \textit{Herschel} guaranteed time key program  
 The Dwarf Galaxy Survey 
 \citep[DGS,][]{RR13}.
The surprising variety of SED shapes found therein \citep[see also HAPLESS results, ][]{Clark:2015}
can be summarized as
(1) a broadening of the IR peak of the SED, implying an additional warmer dust component;
(2) an excess of emission in the submm ($\sim$500 $\mu$m) that causes a flattening of the submm/FIR slope;
and (3) a very low intensity of PAH emission features.

To fit these particular SED patterns, observers change their modeling relative to the standard 
fixed $\beta=2$ in the FIR-submm range \citep[see Eq. \ref{ModBB} and ][]{RR13} 
and/or add new ad hoc additional dust components to it \citep{RR14,RR15,Clark:2015}.
With the aim of going a step further and providing  a physical explanation to these particular
emission features, the SEDs of a sample of 27 simulated star-forming
dwarf galaxies were analyzed in detail and compared to DGS, HAPLESS, and dwarf KINGFISH galaxies.
The sample of simulated dwarf galaxies comes from a single simulation run with 
 the GASOLINE code \citep{wadsley04},
out of initial conditions provided by the  CLUES project
(Constrained Local UniversE Simulations, \citealt{gottloeber10,yepes14}).
These simulated dwarfs have stellar masses, star formation rates at $z=0$,
HI and H$_2$ contents, and metallicities that satisfactorily reproduce observational dwarf galaxy data.

The IR-submm properties of the simulated sample were obtained from their
 SEDs, calculated using the GRASIL-3D radiation transfer code \citep{Dominguez:2014}.
The particular strengths of GRASIL-3D compared to other codes 
  can be summarized as follows: i) the radiative transfer is solved in a grid; ii) it is designed to separately treat the radiative transfer in molecular clouds (MCs) and in the diffuse cirrus component,
 whose dust compositions are different; 
iii) it takes into account that younger stars are associated with denser ISM environments by means of an age-dependent dust-reprocessing of stellar populations;
iv) it includes a detailed non-equilibrium calculation  for dust grains with diameter smaller 
than $a_{flu} \sim 250$ \AA, as required.
In GRASIL-3D modeling, by MCs we mean the densest subvolumes of the gas density distribution that surround young stars and absorb their high-energy photons. 
These  emissions  are reemitted in the IR-submm spectral range, with the radiative transfer
equation solved with techniques appropriate for thick media \citep{Granato:1994}.

Thanks to the separate treatment of the radiative transfer in MCs and in the cirrus, the IR-submm SEDs
of the simulated CLUES dwarfs  can be  directly decomposed into their MC and cirrus
contributions. It has been found that these two components have 
clearly separated emission maxima,
 with that of the cirrus
 spanning some 50 $\mu$m around $\lambda = 150\,\mu$m, 
a range where the maxima of normal galaxies often lie,
 while that of MCs appears at lower $\lambda$ 
values and with a variety of shapes. The respective maxima imply that
dust grains in MCs reach higher global effective temperatures than their counterparts in the cirrus.
A second important element causing the diversity of emission features is the relative intensities of the MC and the cirrus emission at their maxima. 

\begin{figure*}[h]
     \centering
    {\includegraphics[width= 16 cm]{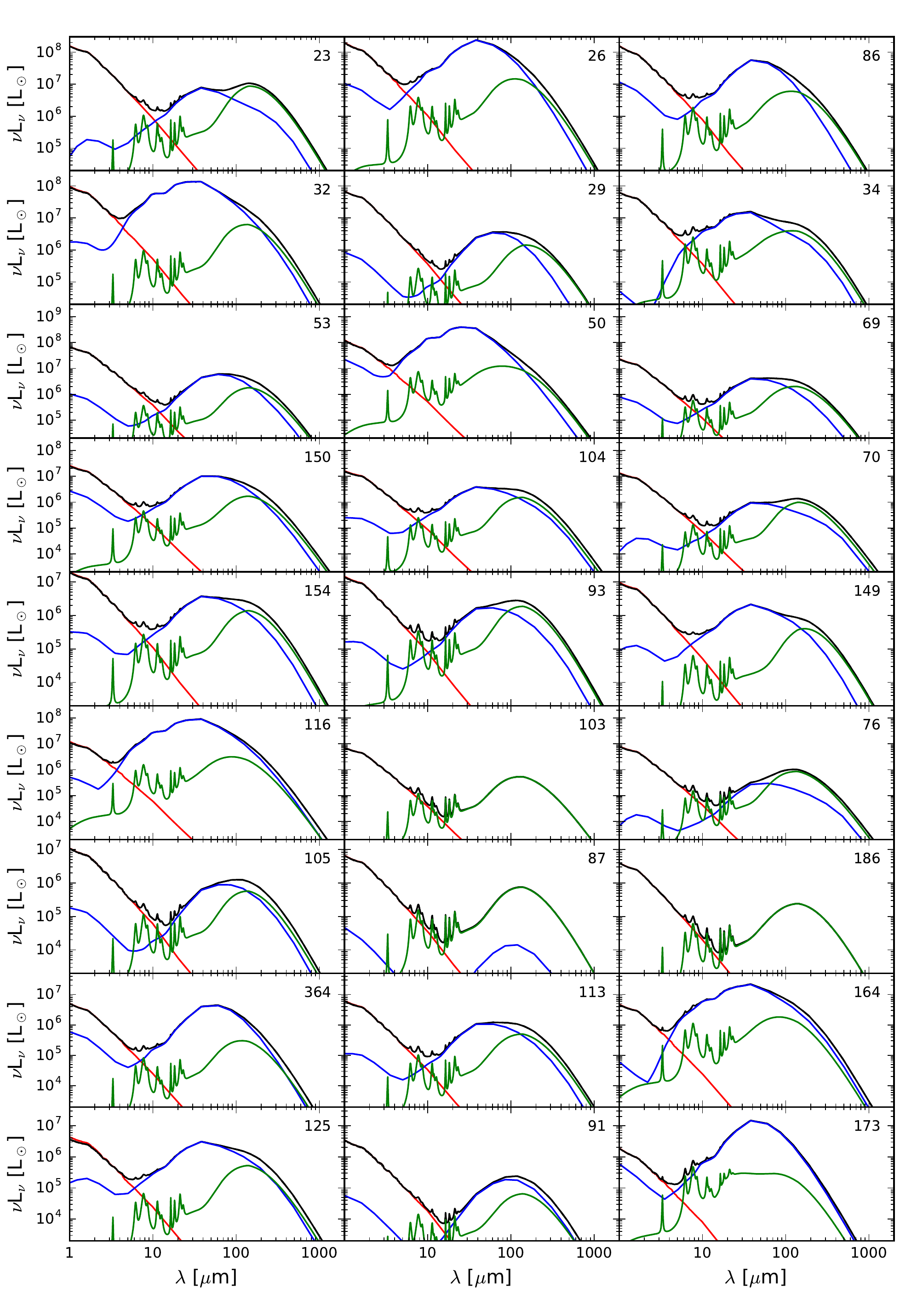}}
   \caption{SEDs of the CLUES sample of local star-forming dwarf galaxies ordered by decreasing stellar mass (from left to right and top to bottom). Red: intrinsic stellar emission when dust is not taken into account. Blue: molecular cloud dust emission. Green: cirrus dust emission. Black: final angle-averaged total emission. 
   }
              \label{FigSEDs}%
    \end{figure*}

\begin{figure*}[h]
      \resizebox{\hsize}{!}{\includegraphics{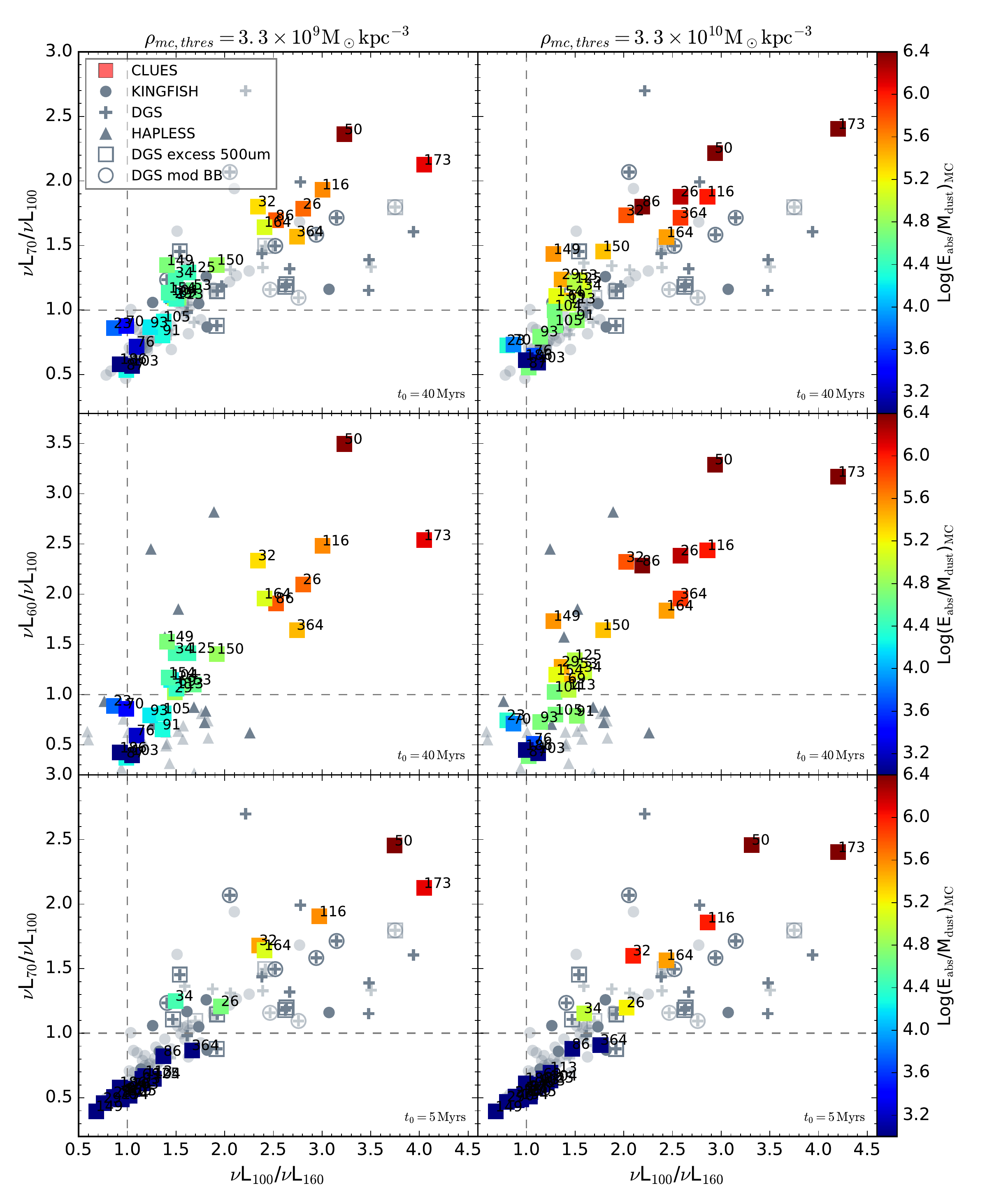}}
   \caption{
   PACS/PACS color-color diagram: $\nu$L$_{70}$/$\nu$L$_{100}$ vs $\nu$L$_{100}$/$\nu$L$_{160}$,  
   which traces the peak of the SED.  CLUES star-forming dwarf galaxies are shown as squares colored by the amount of energy absorbed by molecular clouds per unit dust mass in molecular clouds. Observational data are presented as gray symbols: DGS (crosses), KINGFISH (circles), and HAPLESS (triangles); where faint symbols mark galaxies with
M$_{\rm star}$>10$^9$M$_\odot$.
DGS galaxies whose SED data were fit in \citet{RR15} with an additional emission component as a MIR modified black body are marked with open circles.
DGS galaxies reported in \citet{RR13}  to have an excess of emission at 500 $\mu$m are marked with open squares.
The upper and middle panels show results using a $t_0$=40 Myr molecular cloud escape timescale, while in the lower panels $t_0$=5 Myr.  The left and right columns display the products of adopting a molecular cloud density threshold of $\rho_{mc,thres}$=3.3$\times$10$^{9}$  M$_\odot$kpc$^{-3}$ and $\rho_{mc,thres}$=3.3$\times$10$^{10}$ M$_\odot$kpc$^{-3}$, respectively.
Note that the middle panels show L$_{60}$ instead of L$_{70}$.   
   }
              \label{FigBroad}%
    \end{figure*}

\begin{figure*}  [h]  
     \resizebox{\hsize}{!}{\includegraphics{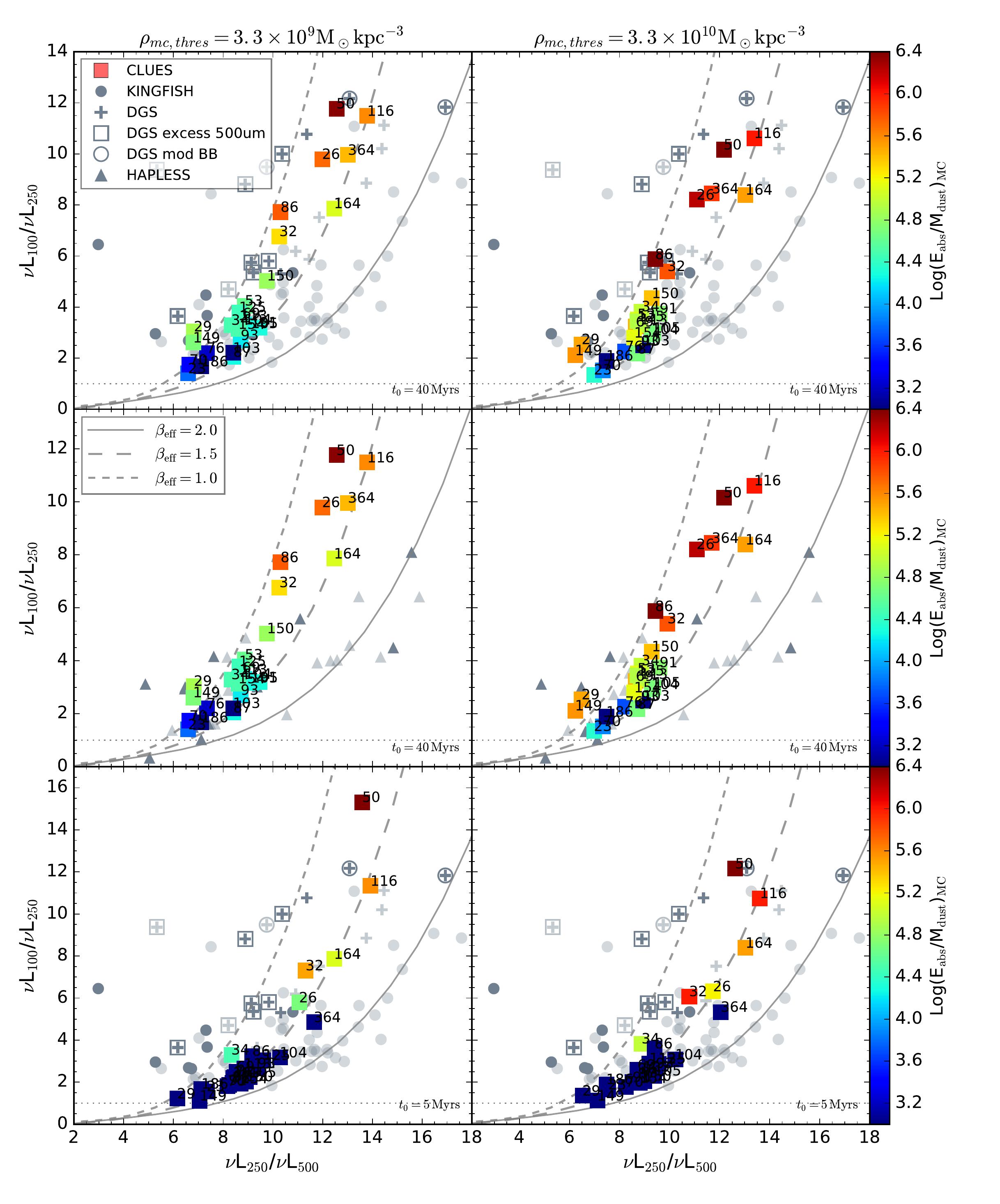}}
   \caption{
   PACS/SPIRE color-color diagram: $\nu$L$_{100}$/$\nu$L$_{250}$ vs $\nu$L$_{250}$/$\nu$L$_{500}$, which traces the variations of the emissivity index $\beta$. Panel distribution, symbols, and color-coding are the same as in Fig. \ref{FigBroad}.
   The curves give the theoretical \textit{Herschel} luminosity ratios for simulated modified black bodies of fixed $\beta_{\rm eff}=$1.0, 1.5, and 2.0 \citep[drawn from Fig. 10 in][]{RR13}.
    }
              \label{FigExcess}%
    \end{figure*}

Two particular situations arise when i) one component (generally MC emission) clearly dominates or when ii)
both components have similar heights at their respective maxima.
Of course a continuum of possibilities exists in between these two, describing  SEDs in Fig. \ref{FigSEDs}. 
When MC emission dominates,
a broad
maximum typically appears
 in the  20 - 40 $\mu$m range,
 providing MIR photons, and (partially) hiding PAH emissions in some cases.
This can explain features 1 and 3 described above. 
Although less common, there are also some examples where cirrus emission dominates the SED and the characteristics of more massive metal-rich galaxies are recovered. 
A different situation arises when both the MC and cirrus emissions have similar intensities at their
(separated) maxima, typically at $\lambda \sim 40$ and $\sim 160\,\mu$m, respectively.
 In this case, their combined emission  is rather flat between the two maxima,
and PAH emission hiding is less likely.

To quantify these behaviors,  an analysis of the PACS/PACS 
  $\nu$L$_{70}$/$\nu$L$_{100}$ vs $\nu$L$_{100}$/$\nu$L$_{160}$
color-color diagram, which traces the peak of the SED,
was performed. Simulated galaxies with SEDs  belonging to the previous situations appear
segregated in this plot, with intermediate situations at intermediate plot locations.
In addition, we found that the driving parameter that causes the segregation
is the amount of energy young stars inject into MCs 
per unit dust mass
$(E_{\rm abs}/M_{\rm dust})_{\rm MC}$,
which takes higher values in the case of dominant MC emission (i.e., a broader IR peak).

Some simulated dwarfs show a flattening of the slope of their SED at submm wavelengths, as compared to that of more massive and metal-rich galaxies.
These simulated dwarfs are precisely those whose
 MC and cirrus contributions
are more separated from each other, 
their combination 
giving an effective excess of emission in the submm region (point 2 above).
It is important to remark that
although we used a fixed value of $\beta = 2$ 
 to model  the individual MC and cirrus emissions,
 a lower final effective slope, $\beta_{\rm eff}$, appears when they blend.

To further explore  this effect, 
we  used
the PACS/SPIRE 
 $\nu$L$_{100}$/$\nu$L$_{250}$ vs $\nu$L$_{250}$/$\nu$L$_{500}$ 
color-color diagram,
which best reflects  the variations of
 $\beta_{\rm eff}$.
By comparing our results to 
 the theoretical luminosity ratios that would be obtained assuming a modified black body of different fixed $\beta_{\rm eff}$ \citep[Fig. 10,][]{RR13},
we found that 
CLUES dwarfs
 lie in between the $\beta_{\rm eff}$ = 1.5 and 1.0 lines, and 
 off
the $\beta_{\rm eff}$ = 2 line. 
A clear trend with the 
$(E_{\rm abs}/M_{\rm dust})_{\rm MC}$
ratio
 is visible, with higher ratio dwarfs lying farther away from the
$\beta_{\rm eff}$ = 2 line (i.e., representing no submm excess) than those with lower ratio.
This is a strong suggestion that the $(E_{\rm abs}/M_{\rm dust})_{\rm MC}$ parameter
adequately describes and drives the submm excess shown by some CLUES dwarfs
 and, presumably, that of DGS, HAPLESS, and dwarf KINGFISH galaxies. 
 This suggestion is corroborated by Fig. \ref{Broad-Corr}.

As said above, a high $(E_{\rm abs}/M_{\rm dust})_{\rm MC}$ ratio implies
MC emission in the MIR range, potentially hiding PAH band emissions.
To be quantitative, we plotted the  $\nu$L$_{8.0}$/$\nu$L$_{24}$ ratios  versus the corresponding stellar masses and 
$(E_{\rm abs}/M_{\rm dust})_{\rm MC}$ ratios.
The result is that
despite the low PAH abundance of the dust model adopted in this study (characterized by a PAH index of $q_{\rm PAH}$=1.12\%),
the lower the $(E_{\rm abs}/M_{\rm dust})_{\rm MC}$ value, the more prominent 
the PAH band emission appears in the SED. This result strongly supports the idea  that within our modeling,
MC dust grains heated by young star light emit in the MIR range and hide the possible residual cirrus PAH band emission.

 GRASIL-3D results depend upon
i) the implementation of the dust-to-gas mass ratio dependence on metallicity, 
ii) the dust model, and 
iii) internal modeling parameters.  
Concerning i),
 a broken power-law dependence of the dust-to-gas mass ratio on metallicity has been set
as suggested by dust evolution models
 in order to better match observed low-metallicity dwarf galaxy dust masses.
  This  implementation does not produce an important effect on the final SEDs with respect to a linear scaling because of the higher influence of high-metallicity grid cells on the final total luminosities.
Regarding ii),
a simple dust model consisting of silicates, graphites, and carbonaceous PAH grains was assumed in both MCs and the cirrus, with the size distributions for each grain type following \citet{Weingartnera:2001} and  \citet{Draine:2007}. 
 The very low PAH band emission observed in dwarf galaxies has compelled modelers to adopt dust models with low PAH abundances. In this study we  also assumed such a model, characterized by a PAH index $q_{\rm PAH}$ of 1.12\%. We showed that although MC dominant emission in dwarf galaxies with very recent star formation can hide the possible PAH emission, which minimizes the ultimate relevance of this parameter, the use of a standard-Galactic value for this index ($q_{\rm PAH}$=4.58\%) would lead to an overly easy visibility of PAH bands in the SEDs of a substantial fraction of the galaxies of our sample, which does not agree with DGS or dwarf KINGFISH statistics.
We also note 
the importance of the prescription used in this work  with respect to the PAH abundance in MCs, where it has been 
drastically lowered  with respect to the cirrus \citep[following][]{Vega:2005} in order to show no PAH band emission.

 iii) 
 Finally,
in order for the SEDs of the simulated  dwarf galaxies to recover the particular features of  DGS, HAPLESS, and dwarf KINGFISH galaxies,  
the GRASIL-3D parameter space had to be somewhat limited. 
In particular, the lack of PAH emission detections demands 
high values for $t_0$,
(the time young stars are enshrouded within MCs)
and
 excludes  low values of the  masses of individual MCs.
The effects on the SED of varying the parameters determining the total MC mass in a given galaxy
($\rho_{mc, thres}$ and $\sigma$) are not that remarkable, 
but their values are limited by the final HI and H$_2$ masses obtained.
Different values for the gas density threshold $\rho_{mc, thres}$ are required 
in order to better match observational molecular gas mass data obtained by means of a 
 metallicity-dependent or a constant-Galactic  CO-to-H$_2$ conversion factor. In particular, a value one order of magnitude higher  is required for the latter case.

To summarize, the analysis of the SEDs of simulated star-forming dwarf galaxies, calculated with 
GRASIL-3D, and their comparison to real data from DGS, HAPLESS, and KINGFISH dwarfs, allows us to conclude the following.
\begin{itemize}
\item The SEDs of simulated dwarf galaxies naturally reproduce the particular spectral features that observed dwarf galaxies show in the IR-submm region.
\item In this spectral range, the SEDs receive two contributions: one from dust in  the diffuse gas component (cirrus),
and a second contribution from dust in the densest component (MCs). 
\item 
The emission maxima of the MC and cirrus components
are clearly separated, with the former typically at $\lambda \sim 40\,\mu$m and the latter at $\lambda \sim 160\,\mu$m.
The intensities of MCs and cirrus at their  maxima can also differ.
\item These two variables (maxima separation and relative intensity of the two dust components at their maxima)
are responsible for shaping the diversity of IR-submm SEDs of dwarf galaxies.
\item More specifically, broad peaks in the MIR region appear when  MCs dominate the total emission,
while SEDs with flat  configurations between  $\lambda \sim 40-160\,\mu$m
 result when the MC and cirrus emission components have similar peak intensities.
\item 
The submm excess can be explained as a result of the combination of the separated MC and cirrus contributions, with no need of imposing a $\beta \neq 2$ value. 
 
\item The lack of detected PAH emission is the result of
i) a low proportion of PAHs in the total dust in the cirrus, plus ii)
 a wealth of MIR emission by hot dust
grains within PAH-devoid MCs, hiding the possible residual cirrus PAH emission. 
\item  The driving parameter adequately describing these particular features is the
amount of energy injected per unit dust mass in MCs. The energy comes from young star emissions.
\item The lower this parameter, the less apparent the particular features.
In the limit of very low parameter values, the higher-mass higher-metallicity galaxy behavior is recovered.
\end{itemize}

 Our conclusions are expected to remain valid when numerical models
of dust formation and evolution are used, for example, if dust formation and evolution
had been self-consistently implemented in the simulation. The reason is that
the relevant regions of the parameter space consistent with observational
data were explored with  enough detail so as  to cover the possible results 
 of numerical models at $z=0$.

Therefore, the GRASIL-3D two-component dust model gives a sound physical interpretation of the emission of low-mass low-metallicity (dwarf)
galaxies, with dust grains within MCs potentially providing a wealth of MIR photons (i.e., the warm  dust
component). Combined with the emissions from colder grains within the cirrus, SEDs in the IR-submm
range are obtained that satisfactorily agree with the particular patterns observed at these wavelengths.

\begin{acknowledgements}

We would like to thank the referee, whose insightful comments
have helped to improve the quality of this paper.
We thank S. Gottloeber and G. Yepes for a careful reading of the manuscript and their comments.
We thank the CLUES collaboration for providing the initial conditions for the simulation analyzed in this work. 
This work was partially supported through MINECO/FEDER (Spain) grants AYA2012-31101  and 
AYA2015-63810-P.
ISS thanks financial support through the first grant.
CB also thanks the Ramon y Cajal program.
We acknowledge the Centro de Computaci\'on Cient\'ifica (Universidad Aut\'onoma de Madrid,
Red Espa\~nola de Supercomputaci\'on) for computational support.
The CLUES simulations were performed and analyzed at the High Performance Computing Center Stuttgart (HLRS).
We thank DEISA for  access to  computing resources  through DECI projects SIMU-LU and SIMUGAL-LU and the generous allocation of resources
from STFC’s DiRAC Facility (COSMOS: Galactic Archaeology),
the DEISA consortium, co-funded through EU FP6
project RI-031513 and the FP7 project RI-222919 (through the
DEISA Extreme Computing Initiative), the PRACE-2IP Project (FP7 RI-283493).

\end{acknowledgements}

%
%

\bibliographystyle{aa}
\bibliography{paper_submm}


\vspace{3cm}

\begin{table*}[p]
\vspace{5cm}
\centering
\caption{General properties of the CLUES star-forming dwarf galaxies.
Stellar mass, total gas mass, HI gas mass, H$_2$ gas mass, total dust mass, star formation rate, and average oxygen metallicity. }
\begin{tabular}{l l l l l l l l}
\hline\hline
Galaxy  & Log \mstar\ & Log M$_{\rm gas}$& Log M$_{\rm HI}$ &  Log M$_{\rm H2}$ & Log M$_{\rm dust}$  & Log SFR$_{\rm z=0}$ & 12+log(O/H)\\
&  (\msun) &  (\msun)&  (\msun)&  (\msun) &  (\msun) & (\msun yr$^{-1}$) & \\
\hline
23	&	8.73	&	8.70	&	8.48	&	7.81	&	5.69	&	-2.85	&	7.62	\\
26	&	8.52	&	8.77	&	8.54	&	7.96	&	5.29	&	-1.55	&	7.32	\\
29	&	8.32	&	8.97	&	8.81	&	7.60	&	5.16	&	-2.56	&	7.01	\\
32	&	8.32	&	8.25	&	7.96	&	7.59	&	5.18	&	-1.79	&	7.82	\\
34	&	8.26	&	8.53	&	8.33	&	7.58	&	5.12	&	-3.39	&	7.21	\\
50	&	7.92	&	8.44	&	8.24	&	7.48	&	4.81	&	-0.76	&	7.25	\\
53	&	8.10	&	8.34	&	8.06	&	7.66	&	4.82	&	-2.15	&	7.52	\\
69	&	7.76	&	8.04	&	7.84	&	7.10	&	4.74	&	-1.89	&	7.60	\\
70	&	7.63	&	7.82	&	7.43	&	7.33	&	4.92	&	-3.18	&	7.69	\\
76	&	7.29	&	7.70	&	7.31	&	7.21	&	4.68	&	-3.60	&	7.75	\\
86	&	8.44	&	8.75	&	8.58	&	7.52	&	5.05	&	-1.06	&	6.98	\\
87	&	7.20	&	7.78	&	7.39	&	7.29	&	4.19	&	-3.18	&	7.45	\\
91	&	6.76	&	7.59	&	7.22	&	7.07	&	3.48	&	-3.33	&	7.17	\\
93	&	7.60	&	7.74	&	7.40	&	7.18	&	4.68	&	-2.78	&	7.89	\\
103	&	7.30	&	7.37	&	7.23	&	0.00	&	4.04	&	-2.89	&	7.64	\\
104	&	7.64	&	7.80	&	7.38	&	7.34	&	4.66	&	-2.88	&	7.69	\\
105	&	7.29	&	7.80	&	7.39	&	7.33	&	4.25	&	-2.95	&	7.42	\\
113	&	7.13	&	7.50	&	7.22	&	6.81	&	4.25	&	-2.80	&	7.76	\\
116	&	7.44	&	7.81	&	7.43	&	7.31	&	4.47	&	-2.63	&	7.65	\\
125	&	6.94	&	7.46	&	7.11	&	6.93	&	4.40	&	-2.87	&	7.69	\\
149	&	7.49	&	7.42	&	7.23	&	6.36	&	4.54	&	-2.78	&	8.00	\\
150	&	7.75	&	8.27	&	7.99	&	7.59	&	4.72	&	-1.66	&	7.35	\\
154	&	7.62	&	7.81	&	7.56	&	7.05	&	4.67	&	-2.38	&	7.72	\\
164	&	7.12	&	7.47	&	7.03	&	7.04	&	4.28	&	-3.37	&	7.66	\\
173	&	6.30	&	7.48	&	7.00	&	7.09	&	3.09	&	-3.11	&	7.08	\\
186	&	7.17	&	7.31	&	7.17	&	0.00	&	3.97	&	-3.20	&	7.60	\\
364	&	7.16	&	7.46	&	7.08	&	6.95	&	3.69	&	-2.79	&	7.50	\\
\hline
\end{tabular}
\label{Table1}
\end{table*}

\end{document}